\documentclass[a4paper,11pt]{JHEP}
\usepackage[T1]{fontenc}
\usepackage{ae,aecompl,aeguill}		
\usepackage[ansinew]{inputenc}
\usepackage{amsmath}
\usepackage{amssymb}
\usepackage{slashed}
\usepackage{graphicx}
\usepackage{subfigure}

\newcommand{\be}{\begin{equation}}
\newcommand{\ee}{\end{equation}}
\newcommand{\beqa}{\begin{eqnarray}}
\newcommand{\eeqa}{\end{eqnarray}}
\newcommand{\bseq}{\begin{subequations}}
\newcommand{\eseq}{\end{subequations}}

\newcommand\m{\mu}

\newcommand\CLASS{{\tt CLASS}}
\newcommand\CAMB{{\tt CAMB}}
\newcommand\CMBFAST{{\tt CMBFAST}}
\newcommand\RECFAST{{\tt RECFAST}}

\newcommand\wi{0.48\textwidth}
\newcommand\g{\gamma}
\newcommand\tc{\tau_c}

\newcommand\di{\mathrm{d}}
\newcommand\s{\sigma}

\newcommand\nur{{\tt ur}}
\newcommand\cdm{{\tt cdm}}

\preprint{CERN-PH-TH/2011-082, LAPTH-010/11} 

\title{The Cosmic Linear Anisotropy Solving System\\ (CLASS) II: Approximation schemes} 

\author{Diego Blas$^a$, Julien Lesgourgues$^{a,b,c}$,
Thomas Tram$^{d,b}$\vspace{.2cm}\\ {$^a$}Institut de
Th\'eorie des Ph\'enom\`enes Physiques,\\ \'Ecole Polytechnique
F\'ed\'erale de Lausanne,\\ CH-1015, Lausanne,
Switzerland.\vspace{.2cm}\\ {$^b$} CERN, Theory Division,\\ CH-1211
Geneva 23, Switzerland.\vspace{.2cm}\\ {$^c$} LAPTh (CNRS -
Universit\'e de Savoie), BP 110,\\ F-74941 Annecy-le-Vieux Cedex,
France.\vspace{.2cm}\\ {$^d$} Department of Physics and Astronomy,\\
University of Aarhus,\\ DK-8000 Aarhus C, Denmark.} 

\abstract{ Boltzmann codes are used extensively by several groups for
  constraining cosmological parameters with Cosmic Microwave
  Background and Large Scale Structure data. This activity is
  computationally expensive, since a typical project requires from
  $10^4$ to $10^5$ Boltzmann code executions.  The newly released code
  \CLASS{} (Cosmic Linear Anisotropy Solving System) incorporates
  improved approximation schemes leading to a simultaneous gain in
  speed and precision. We describe here the three approximations used
  by \CLASS{} for basic $\Lambda$CDM models, namely: a baryon-photon
  tight-coupling approximation which can be set to first order, second
  order or to a compromise between the two; an ultra-relativistic
  fluid approximation which had not been implemented in public
  distributions before; and finally a radiation streaming
  approximation taking reionisation into account.}

\begin{document}

%%%%%%%%%%%%%%%%%%%%%%%%%%
\section{Introduction}
%%%%%%%%%%%%%%%%%%%%%%%%%%

Boltzmann codes have experienced considerable improvements in terms of
precision and speed with respect to the pioneering {\tt COSMICS}
package \cite{Bertschinger:1995er}. In each new public code
(\CMBFAST{} \cite{Seljak:1996is}, \CAMB{} \cite{Lewis:1999bs}, {\tt
CMBEASY} \cite{Doran:2003sy}), several sophisticated optimisation
methods and approximation schemes have been introduced. Efforts on
this side keeps being justified for two reasons. On the one hand, we
need to fit data with higher and higher precision. For instance, the
analysis of Planck data requires much more accurate theoretical
predictions than for WMAP
\cite{Hamann:2009yy}. On the other hand, a growing number of
cosmologists are interested in fitting cosmological data with several
extensions of the minimal cosmological model, in order to probe new
physics. This requires running parameter extraction algorithms on
computer clusters, often involving $10^{4}$ or $10^5$ Boltzmann code
executions (for each new model or new combination of data sets).
Hence, any way to speed up Boltzmann codes without loosing precision
is useful.

In front of such needs, a new code, the Cosmic Linear Anisotropy
Solving System (\CLASS{}) \cite{class_general}, has just been
released\footnote{available at {\tt http://class-code.net}}. The goal
of this project is not just to improve speed and precision, but also
to provide a flexible and user-friendly code that can be easily
generalized to non-minimal cosmological models.  In this paper, we do
not discuss flexibility issues and only concentrate on the improved
approximation schemes used by \CLASS{}, in the strict context of
minimal $\Lambda$CDM cosmology. Extensions requiring extra
approximations may be introduced and discussed case by case in the
future. In a companion paper
\cite{class_ncdm}, we already discuss the approximation specific to
massive neutrinos and non-cold dark matter relics. A comparison
between the power spectra obtained by \CAMB{} and \CLASS{} for the
minimal $\Lambda$CDM model, as well as estimates of the relative speed
of the two codes, is presented in~\cite{class_comp}.

The next three sections describe: a baryon-photon tight-coupling
approximation which can be set to first order, second order or to a
compromise between the two (Sec.~\ref{sec_tca}); an ultra-relativistic
fluid approximation which had not been implemented in public
distributions of Boltzmann codes before (Sec.~\ref{sec_ufa}); and a
radiation streaming approximation consistently including reionisation
(Sec.~\ref{sec_rsa}). Appendix~\ref{sec_stiff} describes the stiff
integrator which can be used by \CLASS{} as an alternative to the
Runge-Kutta integrator: without this integrator, it would have been
essentially impossible to launch test runs with no approximation
schemes, to evaluate the error induced by these schemes; moreover,
this integrator gives better performances even in the presence of
approximations, and is set to be the default integrator in
\CLASS{}. We summarize our full approximation landscape in
Sec.~\ref{sec_summary}.

Note that throughout this paper, when discussing \CAMB{} and
\CLASS{}, we refer to the versions available at the time of preparing
this manuscript, i.e. the January 2011 version of \CAMB{} and {\tt
v1.1} of \CLASS{}.

%%%%%%%%%%%%%%%%%%%%
\section{Tight Coupling Approximation (TCA) \label{sec_tca}}
%%%%%%%%%%%%%%%%%%%%%%

Before recombination, when the opacity $a n_e\sigma_T$ is very large,
the equations governing the tightly coupled baryon-photon fluid form a
stiff system. Indeed, the opacity defines a conformal time scale of
interaction $\tau_c \equiv (a n_e\sigma_T)^{-1}$ considerably smaller
than that on which most of the modes actually evolve, namely $\tau_H =
a/a'$ for super-Hubble scales and $\tau_k = 1/k$ for sub-Hubble scales
(see Fig.~\ref{fig_thermo}). Standard integrators like Runge-Kutta
algorithms would be very inefficient in solving such a system. This
motivated Peebles \& Yu to introduce a simplified system of
differential equations valid in the regime of small $\tau_c/\tau_H$
and $\tau_c / \tau_k$ (tight coupling approximation or TCA)
\cite{Peebles:1970ag}. The overall idea is that quantities which are
vanishingly small in the limit $\tau_c \to 0$ are solved
perturbatively in $\tau_c$, and these analytical expressions are used
in the numerical code solving the remaining differential equations of
the system.

\FIGURE{
%\begin{figure}
%
\includegraphics[width=\wi]{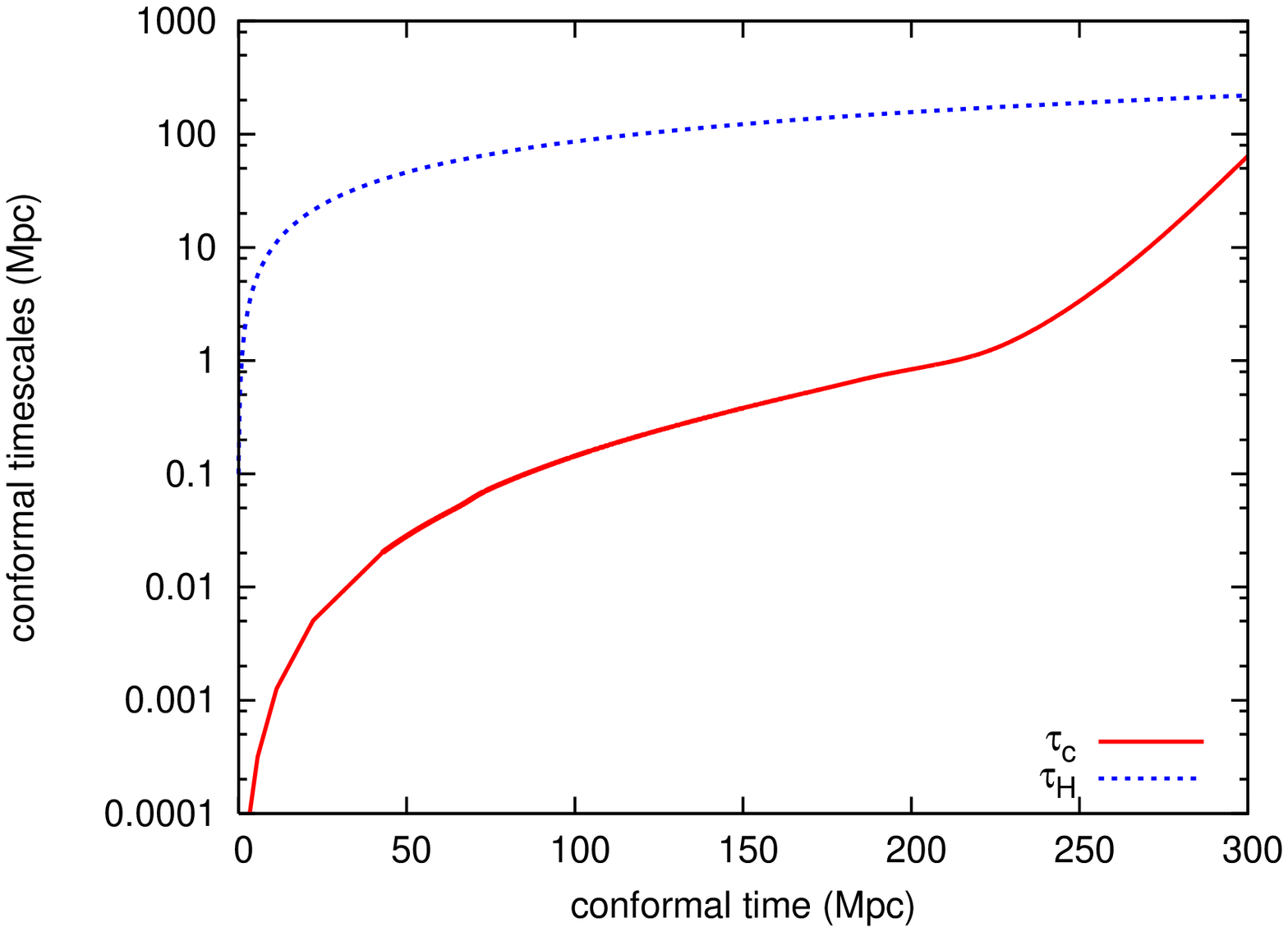}
\includegraphics[width=\wi]{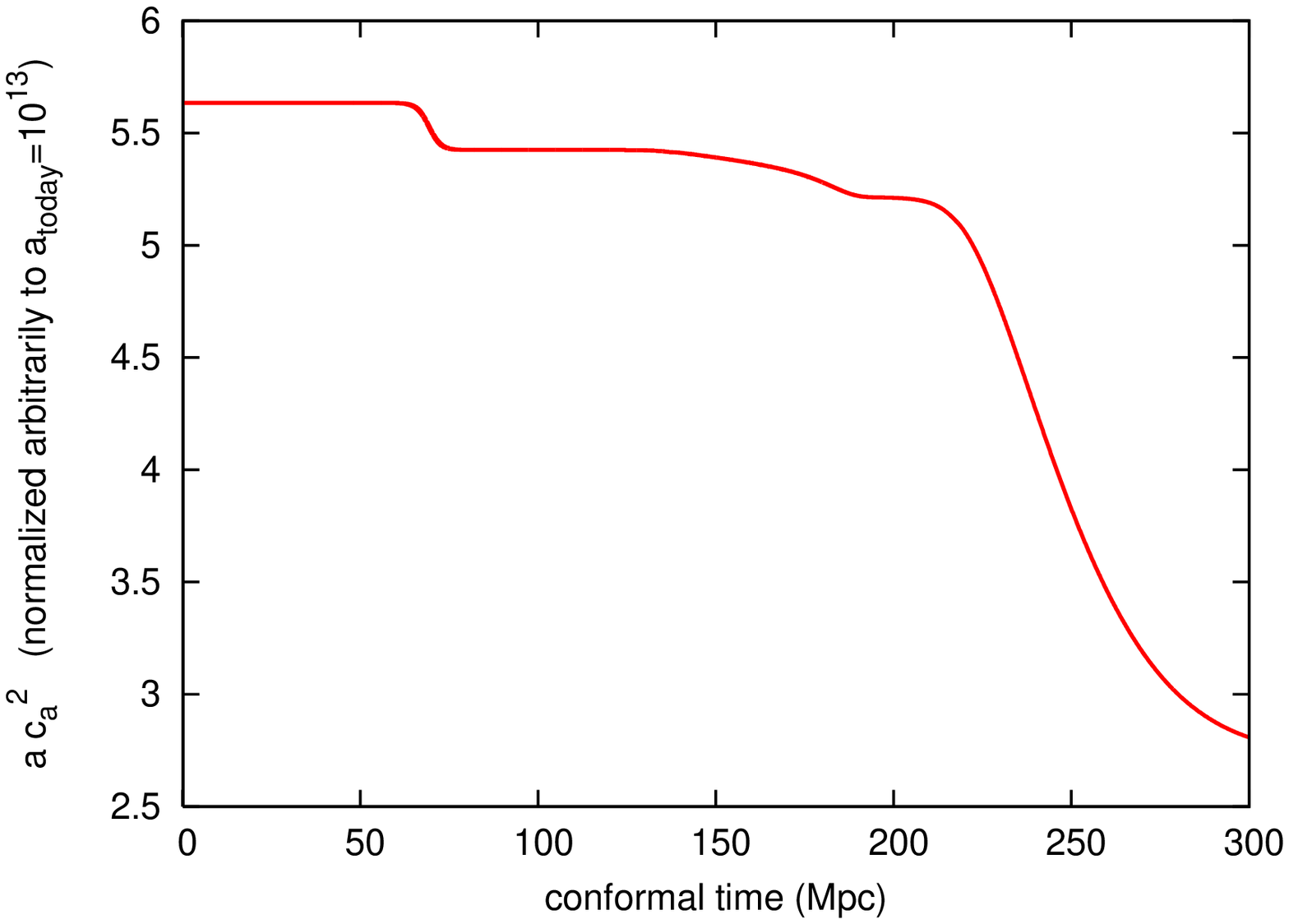}
\caption{\label{fig_thermo} (Left) Evolution characteristic conformal
  time scales in units of Mpc. Before recombination, the baryon-photon
  interaction time scale $\tau_c=(a n_e \sigma_T)^{-1}$ is much
  smaller than the Hubble time scale $\tau_H=a/a'$. For each mode and
  in a given range, it is also smaller than the acoustic oscillation
  time scale $1/k$.  (Right) Evolution of the product $a \, c_a^2$
  scale factor times baryon adiabatic sound speed), with an arbitrary
  normalization of the scale factor, before and during hydrogen
  recombination (in this model, the visibility function peaks at
  $\tau_{rec}=278$~Mpc). For $\tau < 200$~Mpc, helium recombination
  leads to a variation of this product by approximately $10\%$.  }
%\end{figure}
}

The \CLASS{} user can choose between two integrators for the system of
linear perturbations. One of them ({\tt ndf15}, see
Appendix~\ref{sec_stiff}) is optimised for stiff equations, and shows good
performances even in the tight coupling regime. However, by reducing
drastically the number of equations to integrate, any TCA scheme leads
to a speed up even in presence of such an integrator.

Tight coupling equations have already been derived and improved by
many authors after Peebles \& Yu's seminal paper. For instance, the
TCA formulas presented in Ma \& Bertschinger \cite{Ma:1995ey} are
derived to first order in $\tau_c$ (omitting some polarisation terms
which contribute to the photon shear at this order). Lewis et
al. implemented the full first-order solution in \CAMB{}
\cite{Lewis:1999bs}, also relaxing Ma \& Bertschinger's assumption that
$\tau_c \propto a^2$. Doran implemented some improved formulas in {\tt
CMBEASY} which are valid in the Newtonian gauge, and include a few
contributions beyond order one \cite{Doran:2005ep}.

As we will see below, the first-order TCA formulas provide poor
approximations to the baryon-photon differential energy flux and to
the photon shear at large times. This is not much of a problem if one
switches to the exact\footnote{ It should be intended that
  throughout this section, the word ``exact'' is intended in the sense
  ``without using the TCA''. Obviously, our equations are never exact
  since they rely on a number of common and well-justified approximations,
  e.g.: linear perturbations, pressureless CDM, approximate expression
  for photon-baryon coupling, baryon pressure neglected in several
  equations, etc.} equations early enough. However, a better scheme
would allow to switch off the TCA later, and to save a lot of
integration time without loosing precision.

Here, we will derive the full second-order TCA formulas in the
synchronous gauge.  While this work was in preparation, Cyr-Racine and
Sigurdson published a paper on exactly the same topic
\cite{CyrRacine:2010bk}. We will show that the numerical results from
our approach coincide with those of \cite{CyrRacine:2010bk}.  Another
recent paper discussing the TCA beyond first order and its
implementation in second order cosmological perturbation theory is
\cite{Pitrou:2010ai}: this work actually proposes a systematic way to compute
high-order corrections in a given model and at any order\footnote{
As pointed out by the author of \cite{Pitrou:2010ai}, this
method can be implemented numerically provided that the code computing
the evolution of thermodynamical variables outputs fully continuous
and derivable functions of time, which is precisely the case with
the \CLASS{} version of \RECFAST.}.

%%%%%%%%%%%%%%%%%%%%%%%%%%
\subsection{Full equations}
%%%%%%%%%%%%%%%%%%%%%%%%%%

In the following we will adopt the notation of Ma \& Bertschinger
\cite{Ma:1995ey}. The baryon perturbations will be characterized by
the energy density contrast $\delta_b$ and the divergence of the fluid
energy flux $\theta_b$. These quantities satisfy the
equations\footnote{We will work in Fourier space and stick to the
synchronous gauge in conformal time, which we will denote by
$\tau$. We use the prime to denote derivative with respect to
conformal time.}
\bseq
\label{eq:bar}
\begin{eqnarray}
&\delta_b'=-\theta_b-\frac{1}{2}h',\label{deltabdot}\\
&\theta_b'=-{\cal H} \theta_b+c_s^2k^2\delta_b+\frac{R}{\tc} \Theta_{\gamma b},\label{thetabdot}
\end{eqnarray}
\eseq with ${\cal H} = aH = a'/a$, where $a$ is the scale factor, and
we have defined $\Theta_{\gamma b}\equiv\theta_\gamma-\theta_b$ and
$R\equiv\frac{4\bar \rho_\gamma}{3\bar \rho_b}$.  The quantity
$\Theta_{\gamma b}$ represents the divergence of the energy flux of
the photons ($\theta_\gamma$) in the frame comoving with the
baryons. Its time-derivative $\Theta'_{\gamma b}$ is often referred
to as the ``baryon-photon slip''. The fields $h$ and $\eta$ represent
the metric perturbations.  As explained in \cite{Ma:1995ey}, the
  baryon pressure perturbation can be safely neglected in the
  continuity equation, but its Laplacian should be kept in the Euler
  equation (giving raise to the term $c_s^2 k^2 \delta_b$) since it
  affects the evolution of very small wavelengths, smaller than the
  baryon Jeans length. In \cite{Ma:1995ey} and many other references,
  the baryon sound speed is identified to the adiabatic sound speed
  \be c_a^2=\frac{k_B T_b}{\m}\left(1-\frac{1}{3}\frac{\di \ln
    T_b}{\di \ln a}\right), \ee where the evolution of the proton
  temperature is given by \be T_b'=-2 {\cal H} T_b+\frac{2\m
    R}{m_e\tc}\left(T_\g-T_b\right), \ee and $\m$ is the mean
  molecular weight. This approximation has been proved to be
  inaccurate in refs.~\cite{Naoz:2005pd,Lewis:2007zh}, but the
  difference is only important for computing the matter power spectrum
  for $k \gg 10\,h$Mpc$^{-1}$. The current version of \CLASS{} ({\tt
    v1.1}) and \CAMB{} (from January 2011) still relies on the
  $c_s^2=c_a^2$ approximation, while future versions of both codes are
  likely to switch to the actual sound speed calculation, as in the
  {\tt CAMB\_source}\footnote{\tt http://camb.info/sources/}
  code. This issue is irrelevant for the results of this paper, which
  do not involve very large wavelengths. All numerical results below
  have been obtained using $c_a^2$ instead of $c_s^2$, but our
  formulas can adequately describe the large $k$ range, provided that
  a correct sound speed calculation is performed.

Like the adiabatic sound speed, the actual
baryon sound speed is expected to decrease approximately as
$a^{-1}$, except during helium and hydrogen recombination (see
Fig.~\ref{fig_thermo}).
    
    The characterization of the photon distribution requires the
    determination of its different multipoles $\delta_\gamma$,
    $\theta_\gamma$, $\sigma_\gamma$ and $F_{\gamma l}$ for $l\geq
    3$. These satisfy the recursive Boltzmann equations (Eqs. (63) in
    \cite{Ma:1995ey}):
\bseq
\label{eq:photon}
\begin{eqnarray}
&\delta_\g'=-\frac{4}{3}\theta_\g-\frac{2}{3}h', \label{deltagdot}\\
&\theta_\g'=k^2\left(\frac{1}{4}\delta_\g-\sigma_\g\right)-\frac{\Theta_{\g b}}{\tc},\label{thetagdot}\\
&2\sigma_\g'=\frac{8}{15}\theta_\g-\frac{3}{5} k F_{\g 3}+\frac{4}{15}(h'+6\eta')-\frac{9}{5\tc}\sigma_\g+\frac{1}{10\tc}\left(G_{\g 0}+G_{\g 2}\right),\label{sheargdot}\\
\label{eq:F}&F_{\g l}'=\frac{k}{2l+1}\left[l F_{\g( l-1)}-(l+1)F_{\g (l+1)}\right]-\frac{1}{\tc}F_{\g l},\quad l\geq 3\\
\label{eq:G}
&G_{\g l}'=\frac{k}{2l+1}\left[l G_{\g (l-1)}-(l+1)G_{\g (l+1)}\right]\nonumber\\
&\hspace{4cm}+\frac{1}{\tc}
\left[-G_{\g l}+\frac{1}{2}\left(F_{\g 2}+G_{\g 0}+G_{\g 2}\right)\left(\delta_{l0}+\frac{\delta_{l2}}{5}\right)\right],
\end{eqnarray}
 \eseq
where we remind the reader that $F_{\g 2}=2\sigma_{\g}$. Finally, the previous hierarchical equations are truncated 
 at some $l=l_\text{max}$ following
equation (65) in \cite{Ma:1995ey}, \be F_{\g l_\text{max}}'=k
F_{\g(l_\text{max}-1)}-\frac{l_\text{max}+1}{\tau} F_{\g l_\text{max}}-\tc^{-1}F_{\g
  l_\text{max}}.  
  \ee

%%%%%%%%%%%%%%%%%%%%%%%%%%%%%
\subsection{TCA equations}
%%%%%%%%%%%%%%%%%%%%%%%%%%%%%

From Eqs.~(\ref{eq:bar}) and (\ref{eq:photon}) one sees that the
different time scales in the problem are $\tau_c$, $k^{-1}$ and the
time scale of cosmological evolution $\mathcal H$.  As we will show
explicitly in the next section, it is possible to find a solution
(${\Theta}_{\gamma b}^{'\rm tca}$, ${\sigma}_{\gamma}^{\rm tca}$) for the
baryon-photon slip and the photon shear in terms of $\delta_b$,
$\delta_\gamma$, $\theta_b$ and $\theta_\gamma$, which is valid to any
desired order in the small parameter $\tau_c$.  The knowledge of this
solution helps to reduce the full system of equations ~(\ref{eq:bar})
and (\ref{eq:photon}) to just four of them for the low multipoles of the
distributions. More concretely, one may use Eq.~(\ref{deltabdot}) for
${\delta_b}'$, and Eq.~(\ref{deltagdot}) to determine ${\delta_\g}'$;
the energy fluxes are characterized by the linear combination of
Eqs.~(\ref{thetabdot}) and (\ref{thetagdot}) in which the coupling
term vanishes:
\begin{equation}
{\theta_b}' + R {\theta_\g}' = -{\cal H}\theta_b+c_s^2 k^2 \delta_b+Rk^2\left(\frac{1}{4}\delta_\g-{\sigma}^{\rm tca}_\g\right)~,\label{nocoupling}
\end{equation}
and finally
${\theta_\gamma}' - {\theta_b}'={\Theta}_{\gamma b}^{'\rm tca}$.  
As
desired, this scheme allows us to get rid of any coefficient\footnote{The scale $\tau_c$ appears
now  as a small perturbation.} in $\tau_c^{-1}$.
For practical reasons it is customary to combine linearly the last two equations
in order to eliminate ${\theta_\gamma}'$ and get an expression for 
${\theta_b}'$ only; then ${\theta_\gamma}'$ can  be found from equation
(\ref{nocoupling}). In summary, once ${\Theta}_{\gamma b}^{'\rm tca}$ and
${\sigma}_{\gamma}^{\rm tca}$ are known the goal is to solve the closed system 
formed by the four equations (\ref{deltabdot}), (\ref{deltagdot}), and
\bseq
\begin{eqnarray}
\label{eq:theb}
{\theta_b}' &=& - \frac{1}{(1+R)}\left({\cal H}\theta_b-c_s^2 k^2\delta_b-k^2R\left(\frac{1}{4}\delta_\g-{\sigma}_\g^{\rm tca}\right)+R{\Theta}_{\g b}^{' \rm tca} \right)\label{thetabdot2}~, \\
{\theta_\g}' &=& -R^{-1}\left( \theta'_b+{\cal H}\theta_b-c_s^2 k^2 \delta_b\right)+k^2\left(\frac{1}{4}\delta_\g-{\sigma}_\g^{\rm tca}\right)~. \label{thetagdot2}
\end{eqnarray}
\eseq

%%%%%%%%%%%%%%%%%%%%%%%%%%%
\subsection{Perturbative expansion}
%%%%%%%%%%%%%%%%%%%%%%%%%%%

The aim of this section is to find expressions for $\Theta_{\g b}'$
and $\sigma_\g$ valid at the $n^{th}$ order in $\tc$ (the zero
order is trivial: both species behave as a single perfect fluid, so
that $\Theta_{\g b}$ and all multipoles of the photons beyond
$\delta_\g$ and $\theta_\g$ vanish; the first order can be found in 
\cite{Ma:1995ey}).
We first multiply equations (\ref{thetagdot}) and (\ref{thetabdot}) 
by $\tau_c$:
\bseq
\begin{eqnarray}
\tc\left[{\theta_\g}'-k^2\left(\frac{1}{4}\delta_\gamma-\sigma_\g\right)\right]+\Theta_{\g b}&=&0~,\\
\tc\left[-{\theta_b}' - {\cal H} \theta_b +c_s^2 k^2 \delta_b \right] + R \Theta_{\g b}&=& 0~.
\end{eqnarray}
\eseq
To obtain a differential equation for
$\Theta_{\g b}$ we can combine  the above two equations into
\be
\label{gb}
{\tau_c} \left[{\Theta_{\g b}}'-{\cal H}\theta_b+k^2\left(c_s^2\delta_b-\frac{1}{4}\delta_\gamma+\sigma_\gamma\right)\right]+(1+R)\Theta_{\g b}=0~.
\ee
This equation involves the photon shear, given by the following equation (see (\ref{sheargdot})):
\be
\label{sigma}
\sigma_\g=\frac{\tc}{9}\left[\frac{8}{3}\theta_\g+\frac{4}{3}h'+{8}\eta'-10{\s_\g}'-3
  k F_{\g 3}\right]+\frac{1}{18}\left(G_{\g 0}+G_{\g 2}\right). 
  \ee

Until now, all these equations are exact. In the limit of interest, both of them can be schematically 
written as\footnote{This is immediate for (\ref{gb}). For (\ref{sigma}) if follows from (\ref{eq:G}), and we will
explicitly verify it shortly.}
\be 
\epsilon y(t)'+y(t)/f(t)+\epsilon g(t)=0,
\ee 
where
$\epsilon$ is a small parameter. In our case, $\epsilon$ can be chosen to be 
$\bar\tau_c$, the opacity at an arbitrary time around which the expansion is performed\footnote{In fact, the
small dimensionless parameters will be  $\bar\tau_c  k$ and $\bar\tau_c\bar {\mathcal H}$, with $\bar{\mathcal H}$ evaluated
around the same time as $\bar\tc$.}.
The perturbative solution is given by
\be y(t)=\sum_{n=1}\epsilon^n y_n(t), \quad
y_1=-f g, \quad y_{n+1}=- f y_n' ~. \label{pertexp}
\ee
Notice that for functions $f(t)$ and $g(t)$ with smooth time
variations on the scale $\bar{\tau}_c$, the previous is a perfectly
well defined solution.  Finally, the most general solution is found by
adding to the previous particular solutions the solution of the
homogeneous equation \be
\epsilon y(t)'+y(t)/f(t)=0, \quad y= C e^{-1/\epsilon\int f^{-1}\di
  t}~.  \ee Note that in our case $f$ is always positive, which is
enough to make this part of the solution suppressed very fast. Hence, the
relevant solution is given by Eq.~(\ref{pertexp}), which,
after absorbing the small parameter in the function $\tilde f \equiv
\epsilon f$, reads: \be y(t)=\sum_{n=1}\tilde{y}_n(t), \quad \tilde{y}_1=-\tilde f 
g, \quad \tilde{y}_{n+1}=- \tilde f  \tilde{y}_n' ~. \label{pertexp2}\ee
In terms of these functions, the powers of $\tilde f $ and its derivatives represent
the different orders of the approximation.

%%%%%%%%%%%%%%%%%%%%%%%%%%%%%%%%
\subsection{Second-order approximation}
%%%%%%%%%%%%%%%%%%%%%%%%%%%%%%%%

Using the previous expansion in Eq.~\eqref{gb}, the baryon-photon relative velocity
reads at order two:
\be {\Theta}_{\g b}= \tilde f_\Theta \left(-g_\Theta + \tilde f_\Theta' g_\Theta + \tilde f_\Theta g_\Theta'\right)+O(\bar\tau_c^3)~,
\ee
with 
\be 
\tilde f_\Theta =\frac{\tau_c}{1+R}~, \qquad
g_\Theta =
-{\cal H}\theta_b+k^2\left(c_s^2\delta_b-\frac{1}{4}\delta_\gamma+\sigma_\gamma\right)~.
\ee
We still need to differentiate this equation in order to get a
similar approximation for the slip: 
\be {\Theta}_{\g b}'=
\left(\frac{\tilde f_\Theta'}{\tilde f_\Theta}\right){\Theta}_{\g b}+ \tilde f_\Theta \left(-g_\Theta' + \tilde f_\Theta'' g_\Theta + 2 \tilde f_\Theta' g_\Theta' 
+ \tilde f_\Theta g_\Theta''\right)+O(\bar\tau_c^3)~.
\label{slip0}
\ee
There are several ways to organise and simplify the final result.  In
order to write the first-order term in the same form as in the rest of
the literature, we need to express $g_\Theta'$ in a very peculiar way:
\begin{eqnarray} 
g_\Theta' &=&
-{\cal H}\theta_b'-{\cal H}'\theta_b+k^2\left({c_s^2}'\delta_b+c_s^2\delta_b'-\frac{1}{4}\delta_\gamma'+\sigma_\gamma'\right) \nonumber \\
&=&
-2 {\cal H}\theta_b'-({\cal H}'+{\cal H}^2)\theta_b+k^2\left(({\cal H} c_s^2+{c_s^2}')\delta_b+c_s^2\delta_b'-\frac{1}{4}\delta_\gamma'+\sigma_\gamma'\right) + \frac{R{\cal H}}{\tau_c} \Theta_{\g b} \label{gprime} \\
&=&
2 {\cal H}{{\Theta}_{\g b}}'-\frac{a''}{a}\theta_b+k^2\left(- \frac{{\cal H}}{2}\delta_\gamma
+\bar{c}_s^{2}\delta_b+c_s^2\delta_b'-\frac{1}{4}\delta_\gamma'+2{\cal H}\sigma_\gamma+\sigma_\gamma'\right) + \frac{(2+R){\cal H}}{\tau_c} \Theta_{\g b}~.
\nonumber
\end{eqnarray}
Here we defined $\bar{c}_s^{2}\equiv({\cal H} c_s^2+{c_s^2}')$:
this quantity would vanish if the approximation $c_s^2 \propto a^{-1}$
were valid at all times. In the second line, we used
Eq.~(\ref{thetabdot}), while in the third line we used
(\ref{thetagdot}): so these expressions for $g_\Theta'$ are all exact. 

 The first-order approximation for ${\Theta}_{\g b}'$ is obtained by
replacing the first occurrence of $g_\Theta'$ in Eq.~(\ref{slip0})
with the last expression of (\ref{gprime}), in which we neglect the
terms $2 {\cal H} {\Theta}_{\g b}'$ and $(2{\cal H}\sigma_\gamma +
\sigma_\gamma')$ which represent contributions of higher order. The
final result is:
\begin{equation} {\Theta}_{\g b}' 
= \left(\frac{\tau_c'}{\tau_c} - \frac{2{\cal H}}{1+R}\right)\ {\Theta}_{\g
b} - \tilde f_\Theta \left[ -\frac{a''}{a}\theta_b+k^2\left(-
\frac{{\cal H}}{2}\delta_\gamma
+\bar{c}_s^{2}\delta_b+c_s^2\delta_b'-\frac{1}{4}\delta_\gamma'\right)\right]+O(\bar\tau_c^2).
\label{slip1}
\end{equation}
Note that in the previous expression we used the exact relation $R'=-{\cal H}R$.
% and also that the first term in (\ref{slip1}) is the sum of $(\tilde f_\Theta'/\tilde f_\Theta)\Theta_{\gamma b}$ and of
%the last term in $-\tilde f_\Theta g_\Theta'$.  

For the second-order expression for the slip,
we go back to Eq.~(\ref{slip0}). We replace the first occurrence of $g_\Theta'$
by the full expression (\ref{gprime}), assuming that ${\Theta}_{\g
b}'$ and $(2{\cal H}\sigma_\gamma+\sigma_\gamma')$ have been replaced by their
first-order approximation. Finally, we replace $g_\Theta$, $g_\Theta'$ and $g_\Theta''$ in
the last three terms by their zeroth-order approximation.  The final
result can be written in a compact form:
\begin{eqnarray}
{\Theta}_{\g b}' &=&
(1-2{\cal H}\tilde f_\Theta) 
\left\{
\left(\frac{\tau_c'}{\tau_c} - \frac{2{\cal H}}{1+R}\right)\ {\Theta}_{\g
b} - \tilde f_\Theta \left[ -\frac{a''}{a}\theta_b+k^2\left(-
\frac{{\cal H}}{2}\delta_\gamma
+\bar{c}_s^{2}\delta_b+c_s^2\delta_b'-\frac{1}{4}\delta_\gamma'\right)
\right]
\right\} \nonumber \\
&&- \tilde f_\Theta k^2\left(2{\cal H} \sigma_\gamma+\sigma_\gamma' \right)
+ \tilde f_\Theta \left[\tilde f_\Theta'' g_\Theta + 2 \tilde f_\Theta' g_\Theta' + \tilde f_\Theta g_\Theta''\right]+O(\bar\tau_c^3). \label{slip2}
\end{eqnarray}
This formula requires an expression
for the shear valid at order one. However, to solve equations
(\ref{thetabdot},\,\ref{thetagdot}) consistently to second order, we
need the expression for the shear at the corresponding order. This can
be achieved using (\ref{sigma}). To solve this equation let us first
note that the polarisation multipoles $l=0,2$ obey (cf. \eqref{eq:G})
\be
\label{Geq}
\begin{split}
& G'_{\g 0}=-k G_{\g 1}+\tc^{-1}\left[-G_{\g 0}+\s_\g+\frac{1}{2}\left(G_{\g 0}+G_{\g 2}\right)\right],\\
& G'_{\g 2}=\frac{k}{5}\left(2 G_{\g1}- 3G_{\g 3}\right)+\tc^{-1}\left[-G_{\g 2}+\frac{1}{10}\left(2\s_\g+G_{\g 0}+G_{\g 2}\right)\right],
\end{split}
\ee
from which we see that, at first order in $\bar \tau_c$, $G_{\g 2}\sim G_{\g 0}\sim\sigma_\g$.
Thus, it is consistent to consider these multipoles as $O(\bar\tau_c)$, and write the second order solution to
(\ref{sigma}) following (\ref{pertexp2}) as
\be
\label{sigmag}
\begin{split}
\s_\g&=\frac{\tc}{9}\left[\frac{8}{3}\theta_\g+\frac{4}{3} h'+{8} \eta'\right]+\frac{1}{18}\left(G_{\g 0}+G_{\g 2}\right)\\
&-\frac{10\tc}{9}\frac{\di}{\di \tau}\left(\frac{\tc}{9}\left[\frac{8}{3}\theta_\g+\frac{4}{3} h'+
8 \eta'\right]+\frac{1}{18}\left[G_{\g 0}+G_{\g 2}\right]\right)+O(\bar\tc^3),
\end{split}
\ee
where we used the fact that $F_{\g 3}=O(\bar\tc^2)$. Indeed, the high
photon multipoles obey at leading order (cf. \eqref{eq:F})
\be
\nonumber
F_{\g l}=\frac{l\tc k}{2l+1}F_{\g (l-1)}~.
\ee
The final step involves an evaluation of~\eqref{Geq}. Again, from~\eqref{eq:G},
one finds that 
\be
\begin{split}\nonumber
&G_{\g 1}\sim G_{\g 3}\sim O(\bar\tc^2).
\end{split}
\ee
This allows us to find the perturbative solution to \eqref{Geq}
\be
G_{\g 0}=\frac{5\s_\g}{2}-\frac{25}{4}\tc  \s'_\g+O(\bar\tc^3), \quad G_{\g 2}=\frac{\s_\g}{2}-\frac{5}{4}\tc \sigma'_\g+O(\bar\tc^3).
\ee
From the previous expression and (\ref{sigmag}) we find
\be
\s_\g=\frac{8\tc}{45}\left(2\theta_\g+h'+6\eta'\right)+O(\bar\tau_c^2),\label{sigma1}
\ee
which implies
\be
\s_\g'=\frac{8\tc}{45}\left(2\theta'_\g+h''+6\eta''\right)+\frac{8\tc'}{45}\left(2\theta_\g+h'+6\eta'\right)+O(\bar\tau_c^2).
\ee
Finally, the shear at second order is found from (\ref{sigmag}) to be
\be
\label{sigma2}
\s_\g =\frac{8\tc}{45}\left[\left(2\theta_\g+h'+6\eta'\right)\left(1-\frac{11 \tau_c'}{6}\right) -\frac{11 \tau_c}{6}
\left(2\theta_\g'+h''+6\eta''\right)\right]+O(\bar\tau_c^3).
\ee
The last missing items are the zero-order
expressions for $g_\Theta$, $g_\Theta'$
and $g_\Theta''$ appearing in equation (\ref{slip2}). Noticing that
equation (\ref{eq:theb}) implies
\bseq
\begin{eqnarray}
\theta_b' &=& \frac{1}{1+R}\left(-{\cal H} \theta_b + k^2 c_s^2 \delta_b
+ k^2 R \frac{1}{4} \delta_\gamma \right)+O(\bar\tau_c),\nonumber \\
\theta_b'' &=& \frac{1}{1+R}\left((R-1){\cal H} \theta_b' 
-{\cal H}' \theta_b + k^2\left((c_s^2)' \delta_b + c_s^2 \delta_b'
- \frac{R{\cal H}}{4}\delta_\gamma + \frac{R}{4} \delta_\gamma'\right)\right)+O(\bar\tau_c),\nonumber
\end{eqnarray}
\eseq
we can write these last terms as 
\bseq
\label{eq:gcomp}
\begin{eqnarray}
g_\Theta &=& -{\cal H} \theta_b + k^2(c_s^2 \delta_b - \frac{1}{4} \delta_\gamma)+O(\bar\tau_c),  \\
g_\Theta' &=& -{\cal H} \theta_b' -{\cal H}' \theta_b 
+ k^2 \left[(c_s^2)' \delta_b + \left(\frac{1}{3}-c_s^2\right)\left( \theta_b + \frac{1}{2} h' \right)\right] +O(\bar\tau_c),\\
g_\Theta'' &=& -{\cal H} \theta_b'' -2{\cal H}' \theta_b' -{\cal H}'' \theta_b\nonumber \\
&+& k^2 \left[(c_s^2)'' \delta_b 
- 2(c_s^2)' \left( \theta_b + \frac{1}{2} h' \right)
+ \left(\frac{1}{3}-c_s^2\right)\left( \theta_b' + \frac{1}{2} h'' \right)\right]+O(\bar\tau_c). 
\end{eqnarray}
\eseq The derivation given in \cite{CyrRacine:2010bk} follows
different steps, but since it is still a second-order TCA, the results
should be identical under the approximation $c_s^2\propto a^{-1}$ used
in \cite{CyrRacine:2010bk}, at least up to terms of order three or
higher. For the shear, our expressions are indeed exactly
identical. For the slip, there are so many ways to write the result
and so many terms involved that the comparison is not
trivial. However, by coding the two formulas in \CLASS{} and comparing
the evolution of ${\Theta}_{\g b}$ in the two cases, we found that the
two expressions agree very well, since numerically the difference
appears to be at most of order $O(\bar\tc^3)$.

%%%%%%%%%%%%%%%%%%%%%%%%%%%%%%%%%%%%%%
\subsection{Implementation of various schemes in \CLASS{}\label{tca_comp}}
%%%%%%%%%%%%%%%%%%%%%%%%%%%%%%%%%%%%%%

We implemented various TCA schemes in \CLASS{}, which can be chosen by setting
the flag {\tt tight\_coupling\_approximation} to different values.  We
always start integrating the wavenumbers very deep inside the
tightly-coupled regime. Hence, in contrast to \cite{CyrRacine:2010bk},
we do not need to include terms in $\tau_c$ in the initial conditions.
We always use the set of equations (\ref{deltabdot}),
(\ref{deltagdot}), (\ref{thetabdot2}), (\ref{thetagdot2}) with
different expressions for the slip ${\Theta}_{\g b}^{' \rm tca}$ and the
shear ${\sigma}_{\g}^{\rm tca}$:
\begin{enumerate}
\item first-order expressions (\ref{slip1}, \ref{sigma1}), with the
  approximations $\tc\propto a^2$ and $c_s^2\propto a^{-1}$, used e.g.
  in ref. \cite{Ma:1995ey}. This corresponds to the scheme used by \CLASS{} when  the label {\tt
    tight\_coupling\_approximation} is set to {\tt first\_order\_MB}.
\item 
first-order expressions (\ref{slip1}, \ref{sigma1}) with the only approximation  $c_s^2\propto a^{-1}$, like in 
\CAMB{}. This scheme is used by \CLASS{} when the same flag
is set to {\tt first\_order\_CAMB}. 
\item
exact first-order expressions (\ref{slip1}, \ref{sigma1}) when the flag is set to
{\tt first\_order\_CLASS}.
\item
second-order expressions from \cite{CyrRacine:2010bk} when the flag is set to 
{\tt second\_order\_CRS}.
\item
second-order expressions from Eqs.(\ref{slip2}, \ref{sigma2}) for the flag {\tt second\_order\_CLASS}.
\item finally, second-order expression for the shear, but a reduced
  expression for the slip, involving only the leading order-two terms:
\be
\begin{split}
{\Theta}_{\g b}^{' \rm tca} =&
(1-2{\cal H}\tilde f_\Theta)
\left\{
\left(\frac{\tau_c'}{\tau_c} - \frac{2{\cal H}}{1+R}\right)\ {\Theta}_{\g
b} - \tilde f_\Theta \left[ -\frac{a''}{a}\theta_b+k^2\left(-
\frac{{\cal H}}{2}\delta_\gamma
+c_s^2\delta_b'-\frac{1}{4}\delta_\gamma'\right)
\right]
\right\}
\\
&- \tilde f_\Theta k^2\left[2{\cal H} \sigma_\gamma+\sigma_\gamma' -\left(\frac{1}{3}-c_s^2\right)\left( \tilde f_\Theta
\theta_b' + 2 \tilde f_\Theta' \theta_b \right)
\right], \label{compromise}
\end{split}
\ee
This option is taken when the flag is set to {\tt
compromise\_CLASS}, and is chosen to be the default option in
\CLASS{}.  Notice that the last scheme does not have a term $h''$,
which is advantageous from the computational point of view\footnote{In
order to compute $h''$ one should use one more Einstein equation that
in the standard case, and compute the $\delta T_i^i$ component of the
stress-energy tensor, i.e. the pressure perturbation for all species.}.
\end{enumerate}
To justify the compromise scheme, notice that it encapsulates the
leading order in (\ref{eq:gcomp}) for subhorizon modes (note that
$\theta_b$ has an extra momentum dependence with respect to the other
perturbations, and each time derivative adds one more power of $k$ in
this regime). Thus, from Eq.~(\ref{slip2}) we learn that
Eq.~(\ref{compromise}) implements the leading second order correction
for the modes with a big comoving momentum. It is precisely for these
modes that the first order approximation fails first, which
explains the success of the compromise scheme. To check that the
approximation is indeed correct, we implemented this scheme for
several modes $k$ assuming $\Lambda$CDM.  As illustrated in the next
subsection, this scheme is nearly as good as the full second-order
one, being at the same time much more compact and requiring many less
floating point operations.

%%%%%%%%%%%%%%%%%%%%%%%%%%%%%%%
\subsection{Comparison at the level of perturbations}
%%%%%%%%%%%%%%%%%%%%%%%%%%%%%%%

\FIGURE{
%\begin{figure}
%
\includegraphics[width=\wi]{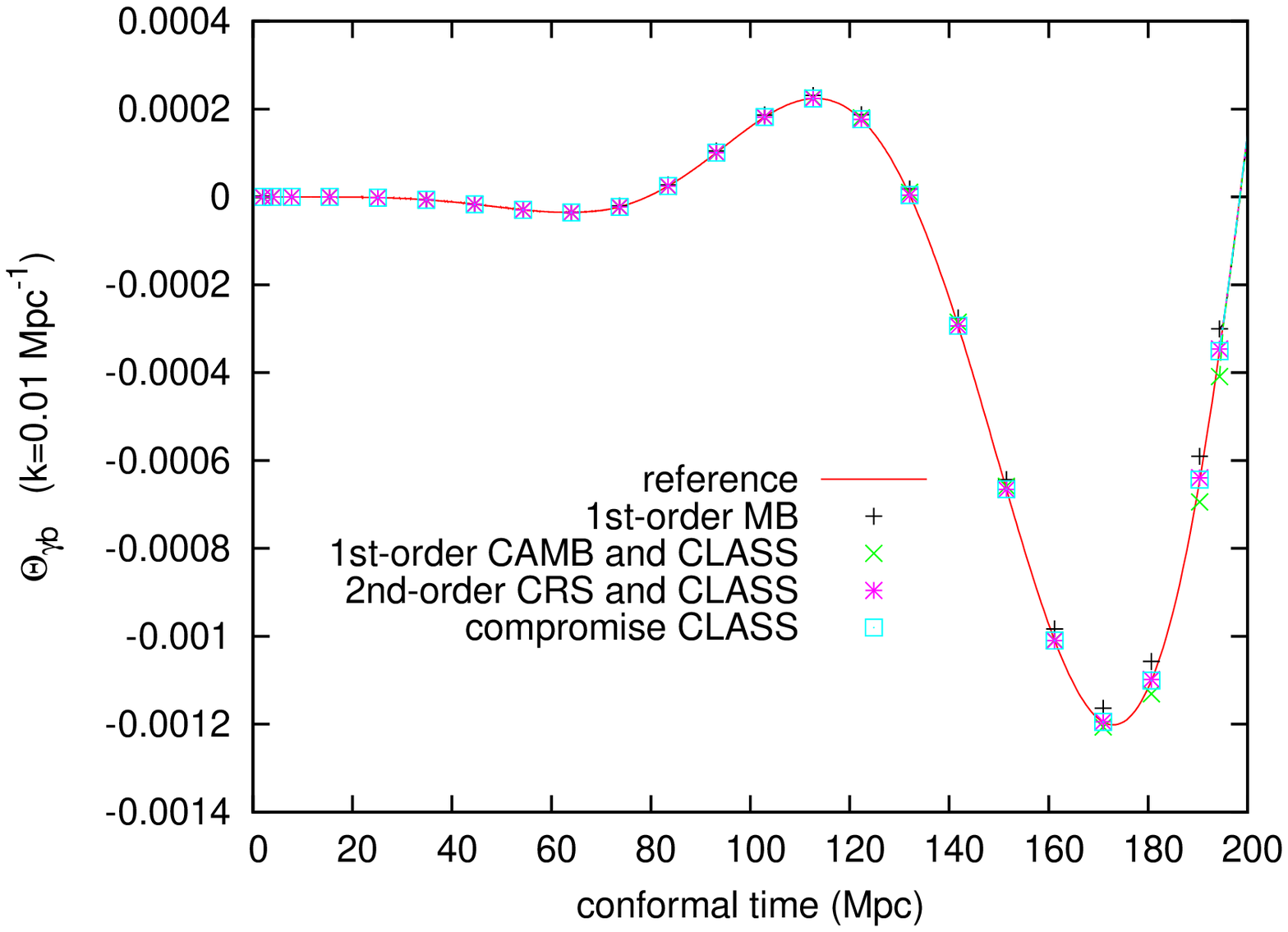}
\includegraphics[width=\wi]{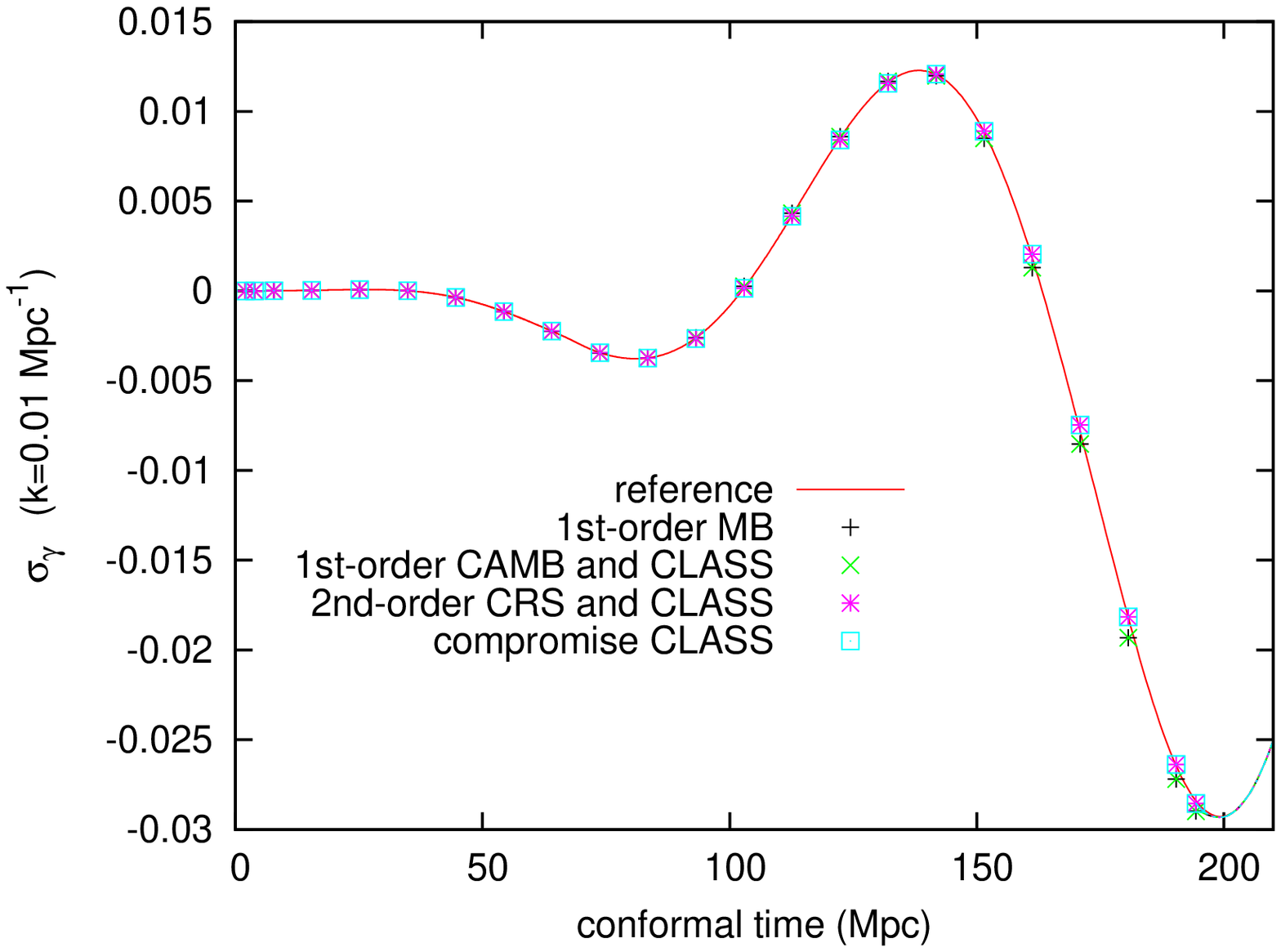}
\caption{\label{tca_var} Evolution of ${\Theta}_{\g b}$ (left) and
  $\sigma_\gamma$ (right) for the mode $k=10^{-2}{\rm Mpc}^{-1}$,
  using the various TCA schemes listed in Sec.~\ref{tca_comp}. In
  each case, the quantities are represented as points when the TCA is
  switched on, and as continuous lines of the same color when exact
  equations take over.  The TCA is switched off at $\tau=1$~Mpc in the
  reference case, and at $\tau=194$~Mpc in all other cases. We show a
  single set of points for cases which are indistinguishable by eye,
  namely: {\tt first\_order\_CAMB} and {\tt first\_order\_CLASS}; and
  also, {\tt second\_order\_CRS} and {\tt second\_order\_CLASS}. The
  default scheme {\tt compromise\_CLASS} is also hardly
  distinguishable from the {\tt second\_order\_CLASS}.  }
%\end{figure}
}

\FIGURE{
%\begin{figure}
%
\includegraphics[width=\wi]{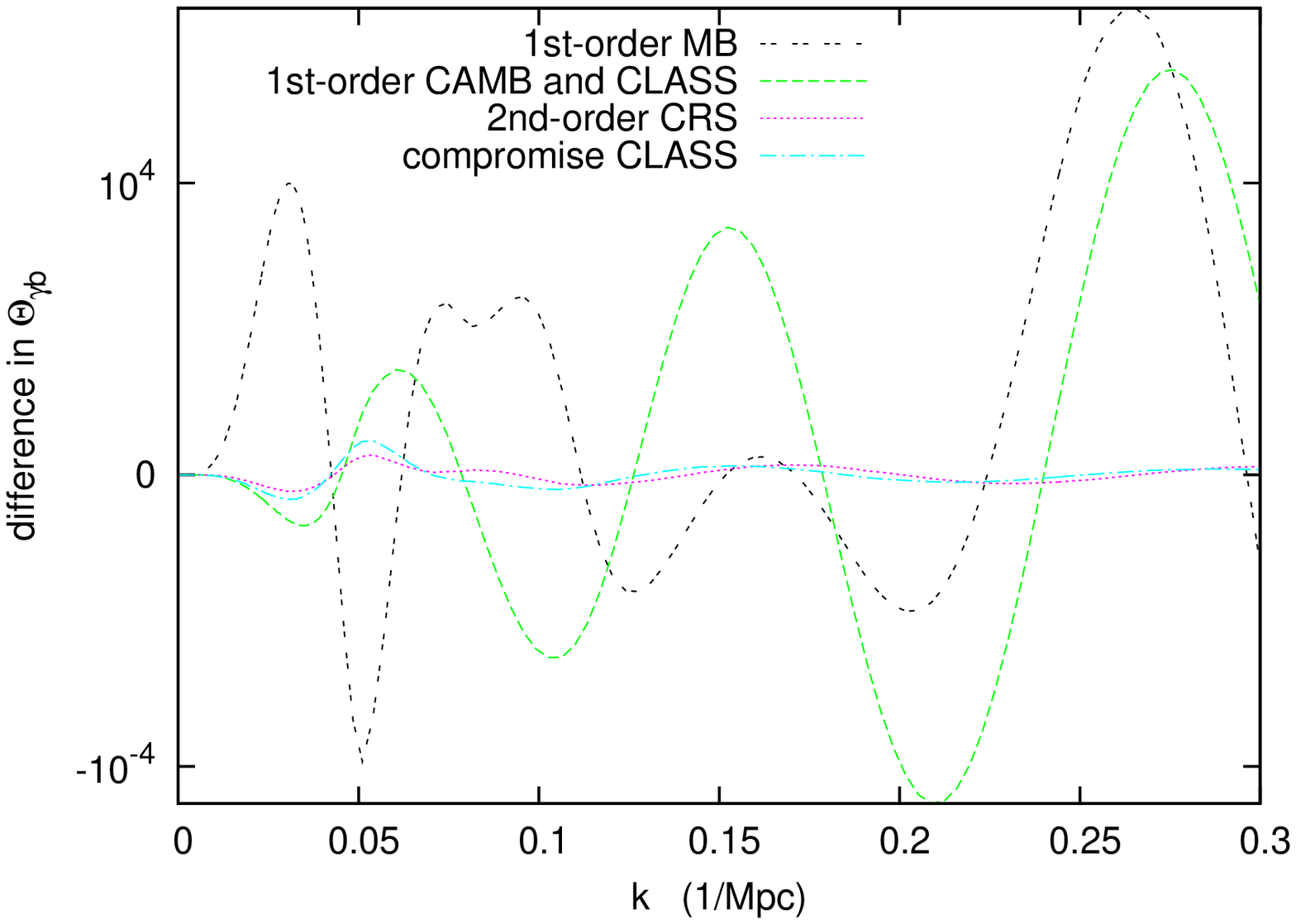}
\includegraphics[width=\wi]{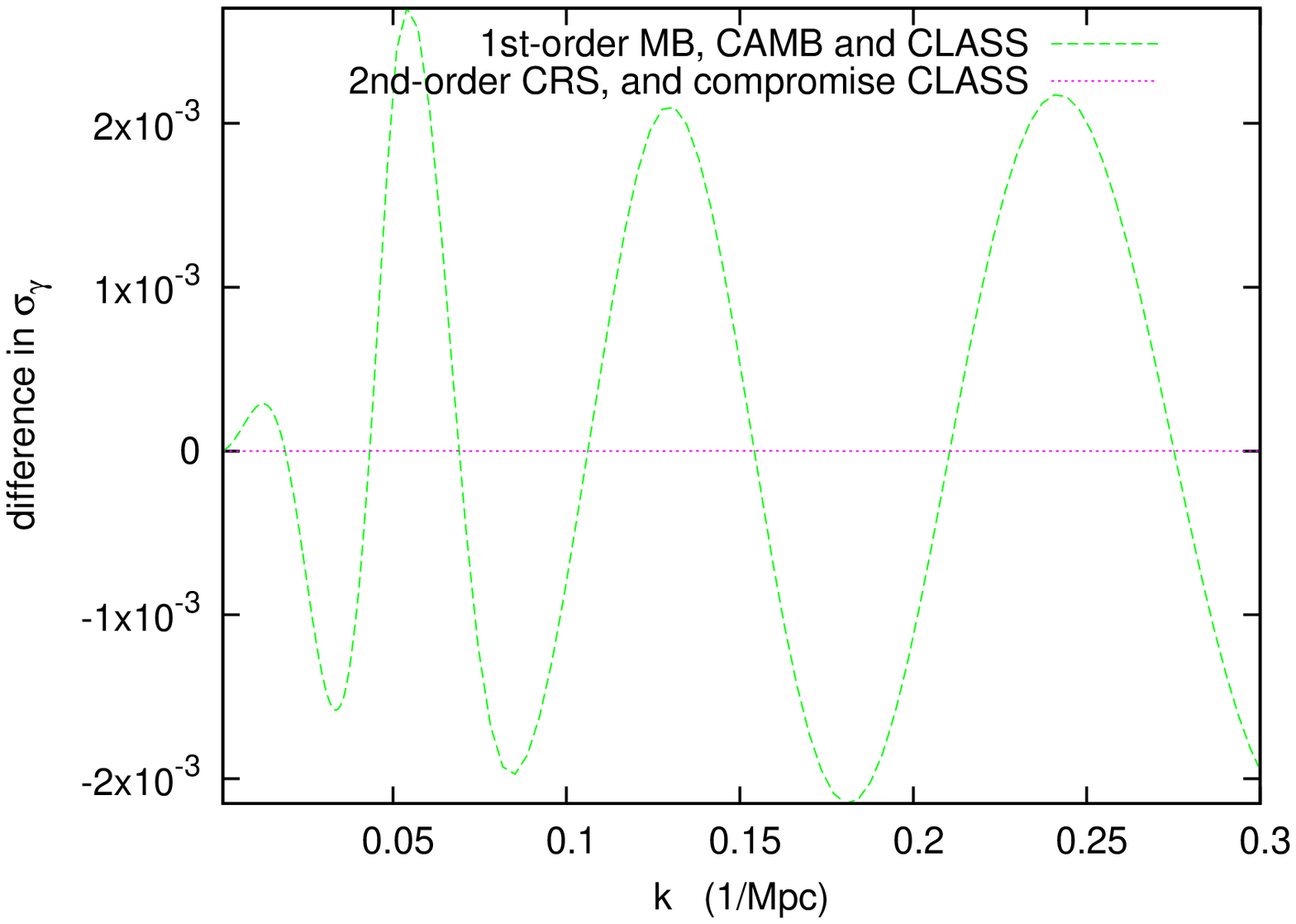}
\caption{\label{tca_snapshot} Comparison of ${\Theta}_{\g b}(k)$ (left) and
  $\sigma_\gamma(k)$ (right) at the time when the TCA is switched off ($\tau=194$~Mpc in this case),
for the different schemes listed in Sec.~\ref{tca_comp}. All cases are compared to the full second-order scheme {\tt second\_order\_CLASS}. For ${\Theta}_{\g b}$, we show a single curve for the cases {\tt first\_order\_CAMB} and {\tt first\_order\_CLASS}, since they are indistinguishable by eye. For $\sigma_\gamma$, all schemes using the first-order shear expression are indistinguishable; this is also true for all cases using the second-order shear expression (despite the fact that  ${\theta}_{\g}$ and metric perturbations are slightly different in each individual case).}
%\end{figure}
}

In Fig.~\ref{tca_var}, we compare these different approximations for a
fixed wave number (chosen to be $k=10^{-2}{\rm Mpc}^{-1}$), and in
Fig.~\ref{tca_snapshot} for a fixed conformal time $\tau$ (chosen to
be the time when the TCA is switched off in the previous example).

Scrutinising first the various first-order schemes, we see a
significant difference at late time between the first two ({\tt MB}
and \CAMB{}), showing that $\tau_c \propto a^{-1}$ is a bad
approximation. However, there is no sizable difference between the
second one (\CAMB{}) and the third one (\CLASS{}) in which the
approximation $c_s^2\propto a^{-1}$ is relaxed. We reach the same
conclusion when comparing second-order schemes with or without the
same approximation. This is not a surprise, since we are only
considering scales larger than the baryonic Jeans length at any
time. When studying very small wavelengths, the \CLASS{} user is free
to choose one of the TCA schemes where the full evolution of $c_s^2$
is automatically taken into account (namely, {\tt first\_order\_CLASS}
or {\tt second\_order\_CLASS}), but as mentioned before, in this
limit, an accurate sound speed computation should also be implemented
in order to relax the $c_s^2=c_a^2$ approximation.

Figs.~\ref{tca_var} and \ref{tca_snapshot} show that all first-order
schemes provide poor approximations for the slip and the shear near
the end of the tightly coupled regime, and hence, inaccurate initial
conditions at the time at which the full equations are turned on. As
expected, second-order schemes work much better.  We find a very good
agreement between {\tt second\_order\_CRS} and {\tt
  second\_order\_CLASS}: this validates both the results of \cite{CyrRacine:2010bk}
and our results. The residual difference is likely due to the fact
that our two independent derivations lead to expressions in which some
higher-order terms (of order $O(\bar\tau_c^3)$) appear in different
ways.

These two schemes also agree to a very good extent with {\tt
  compromise\_CLASS}, which is much more straightforward to code, and
computes the baryon-photon slip with approximately ten times less
operations.  In particular, with this scheme, the code does not even
need to compute the quantities $h''$, $\tau_c''$ and ${\cal H}''$,
which are not so obvious to obtain with few operations and good
accuracy. Hence, this method is set to be the default TCA in \CLASS{}.

%%%%%%%%%%%%%%%%%%%%%%%%%%%%%%%
\subsection{Comparison at the level of temperature/polarisation spectra}
%%%%%%%%%%%%%%%%%%%%%%%%%%%%%%%

First, let us specify which precision parameters in \CLASS{} govern the
evolution of perturbations in the early universe and the TCA switching time: 
\begin{itemize}
\item two parameters define the time at which initial conditions are
  imposed during the tightly-coupled stage. Each wave-number starts being
  integrated (with one of the TCA schemes) as soon as one of the two conditions
  $$(\tau_c/\tau_H) \geq {\tt \tt start\_small\_k\_at\_tau\_c\_over\_tau\_h}$$ or
$$ (\tau_H/\tau_k) \geq {\tt start\_large\_k\_at\_tau\_h\_over\_tau\_k}$$
is fulfilled. The second condition means that at
initial time, wavelengths should be sufficiently far outside the
Hubble scale; the first condition, which over-seeds the second one for
the smallest wave numbers, means that the initial time should not be
too close to recombination.
\item two parameters define the time at
  which the TCA is turned off for each wave number. This happens when one of
  the two conditions $$(\tau_c/\tau_H) \geq {\tt tight\_coupling\_trigger\_tau\_c\_over\_tau\_h}$$ or
  $$(\tau_c/\tau_k) \geq {\tt tight\_coupling\_trigger\_tau\_c\_over\_tau\_k}$$ is fulfilled. \CLASS{} imposes that the TCA switching time should always be chosen after the initial time, which means that the four parameters above should satisfy simple inequalities.
\item one parameter defines the time (common to all wave numbers) at
  which the source functions (leading to the computation of
  temperature and polarisation $C_l$'s) start being sampled and
  stored. This happens when the condition
$$(\tau_c/\tau_H) = {\tt start\_sources\_at\_tau\_c\_over\_tau\_h}$$
is satisfied.
This time can eventually be chosen during
  the tight-coupling regime for the smallest wave numbers.
\end{itemize}

In order to show the impact of these parameters, we take the
set of precision parameters defined in the file {\tt cl\_permille.pre}
of the \CLASS{} public distribution, which corresponds to an accuracy of
at least 0.1\% on each temperature and polarisation $C_l$, and uses
the {\tt compromise\_CLASS} scheme. We then vary the two trigger
parameters mentioned above, as described in Table~\ref{tca_tab}. The
first setting, called {\tt no-tca}, corresponds to switching off the
tight coupling approximation immediately after setting the initial
conditions, so that no TCA is ever used. This leads to reference
temperature/polarisation spectra with respect to which all the other
results of this section are compared. The settings called {\tt tca1},
{\tt tca2} and {\tt tca3} introduce from 0.02\% to 0.08\% of error with
respect to the {\tt no-tca} case, as illustrated in Fig.~\ref{tca_cl1}.

In Fig.~\ref{tca_cl2}, we stick to the precision setting {\tt tca3}
and compare the different TCA schemes. The {\tt first\_order\_CAMB}
and {\tt first\_order\_CLASS} results are indistighuishable,
confirming the fact that the approximation $c_s^2\propto a^{-1}$ is
sufficient in practice.  Our second-order results and those derived
from
\cite{CyrRacine:2010bk} are also in perfect agreement. As expected, the
results from the {\tt compromise\_CLASS} scheme are essentially as
good as the full second-order results, while the first-order results
are roughly ten times less accurate.  We also show on this plot the
error produced by the {\tt first\_order\_CLASS} scheme with {\tt tca1}
precision settings, which is similar to that produced by the {\tt
compromise\_CLASS} scheme with {\tt tca3} precision settings. Hence, in
order to estimate the usefulness of going beyond the first order TCA,
we can compare the performances of the code in these two cases.

In Table~\ref{tca_time}, we compare running times in the {\tt no-tca}
case and in the previous two cases, using either the Runge-Kutta or
{\tt ndf15} integrator. The timings displayed here correspond to the
number of seconds spent by our computer in the perturbation module of
\CLASS, in a non-parallel execution. The {\tt ndf15} integrator is
always better, by a huge amount in the {\tt no-tca} case, or by 20 to
30\% in the other cases. Being unaffected by the issue of integrating
a stiff system, the {\tt ndf15} integrator is not
very sensitive to the choice of TCA scheme, with only a 3\% speed up
when using the compromise scheme instead of first-order
schemes. The Runge-Kutta integrator is more sensitive, with a 9\%
speed-up for the compromise scheme. 

We conclude that \CLASS{} benefits
much more from the implementation of our stiff integrator than from
going beyond the first-order TCA. For some particular models, the user
may wish to stick to the Runge-Kutta integrator, in which case the
{\tt compromise\_CLASS} scheme leads to a sizable speed up.

\TABLE{
%\begin{table}
\begin{tabular}{|l|c|c|c|c|}
\hline
& {\tt no-tca} & {\tt tca1} & {\tt tca2} & {\tt tca3}\\
\hline
{\tt tight\_coupling\_trigger\_tau\_c\_over\_tau\_h} & $4.1\cdot10^{-4}$ & $7\cdot 10^{-3}$  & $8\cdot 10^{-3}$ & $9\cdot 10^{-3}$ \\
{\tt tight\_coupling\_trigger\_tau\_c\_over\_tau\_k} & $6.1\cdot10^{-5}$ & $3\cdot 10^{-2}$ & $5\cdot 10^{-2}$ & $8\cdot 10^{-2}$ \\
\hline
\end{tabular}

\caption{\label{tca_tab} 
  Four settings for the precision parameters governing the time 
  at which the TCA is switched off.
}
%\end{table}
}

\FIGURE{
%\begin{figure}[t]
%
\includegraphics[width=\wi]{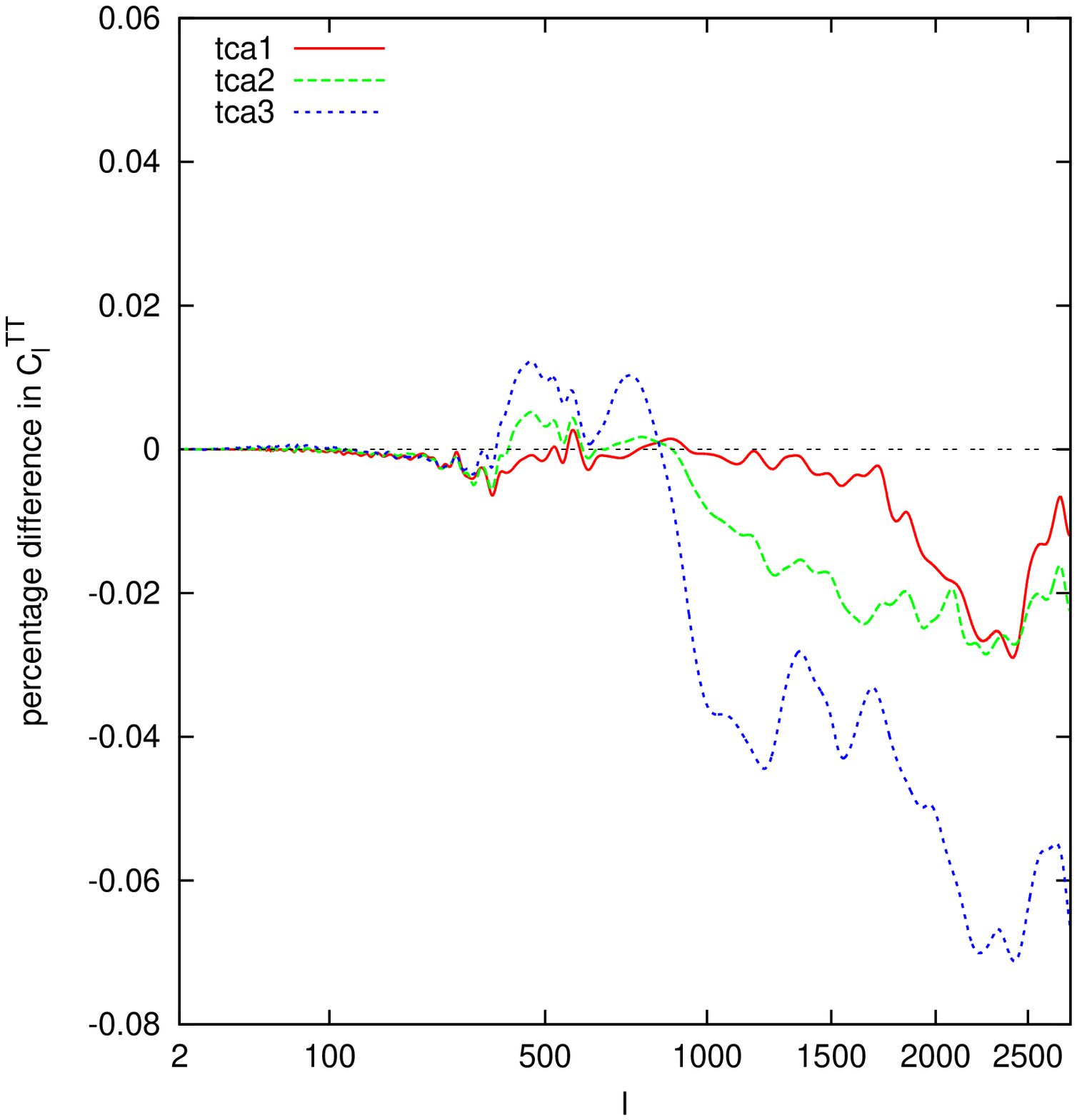}
\includegraphics[width=\wi]{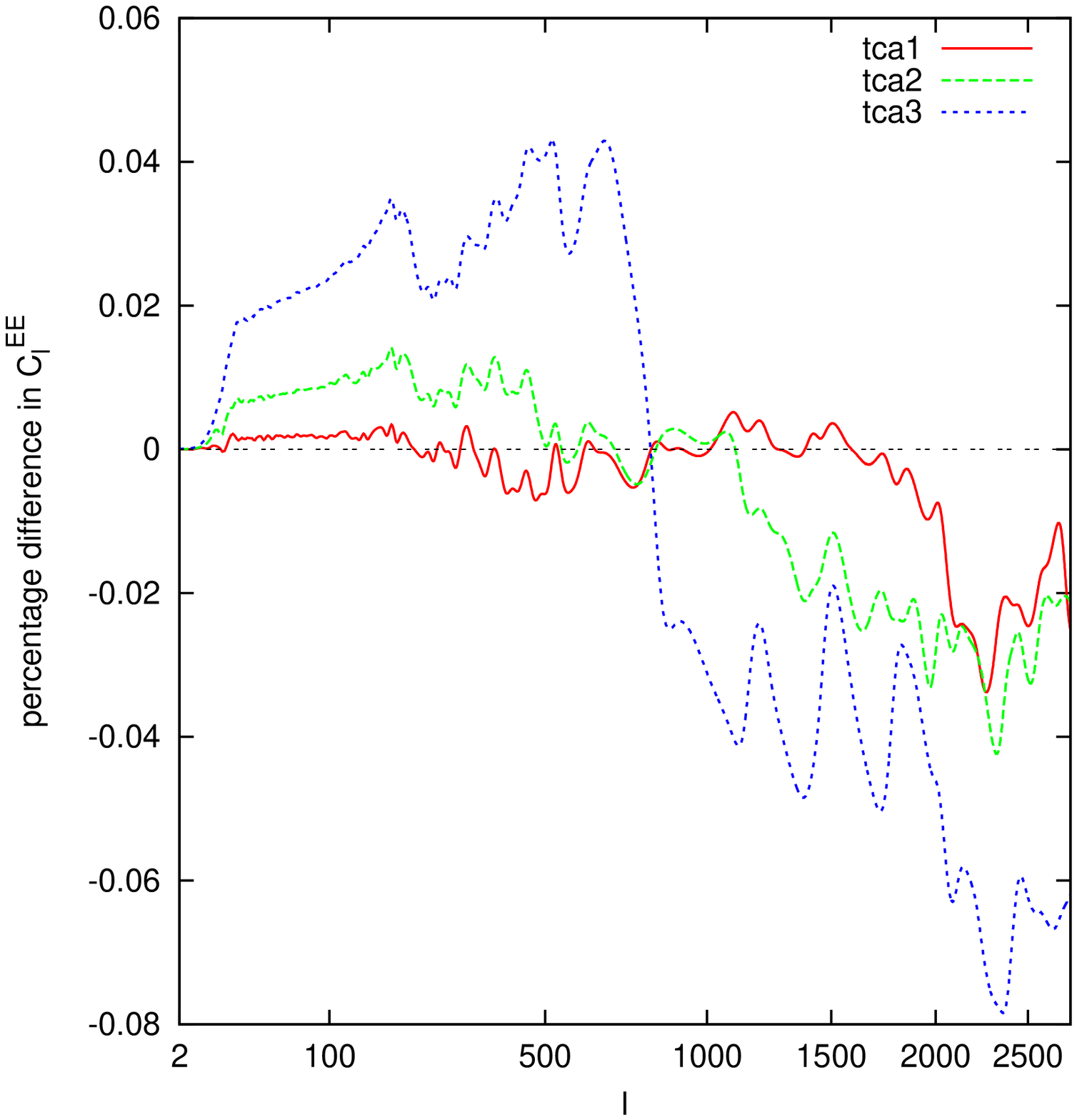}
\caption{\label{tca_cl1} {Impact of precision parameters governing the
    TCA switching time, for temperature (left) and E-polarisation
    (right)}. We show the power spectrum of the three settings {\tt
    tca1}, {\tt tca2} and {\tt tca3} compared to the reference
  spectrum {\tt no-tca} (see Table \ref{tca_tab} for precision
  parameter values).}
%\end{figure}
}

\FIGURE{
%\begin{figure}[h]
%
\includegraphics[width=\wi]{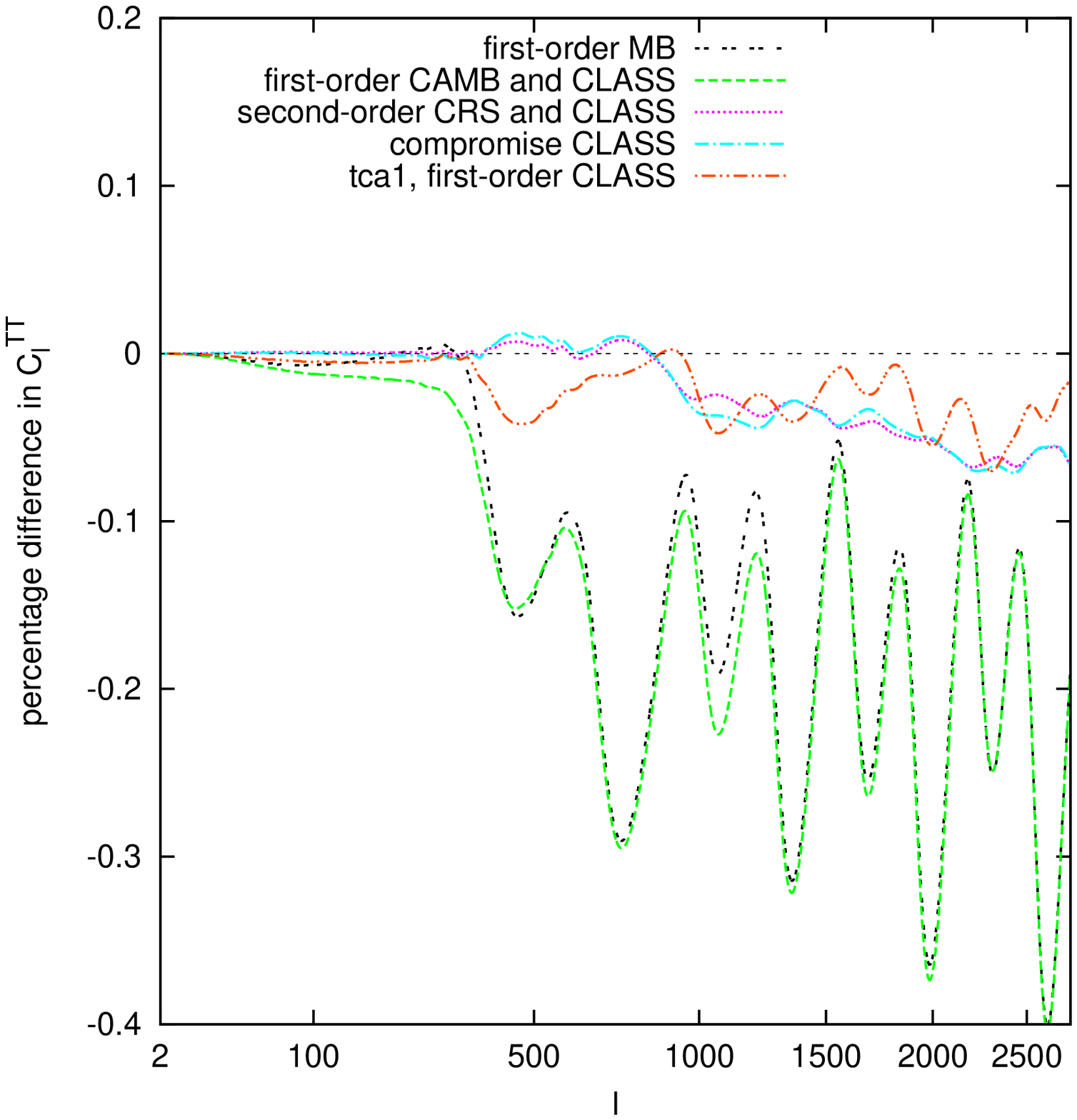}
\includegraphics[width=\wi]{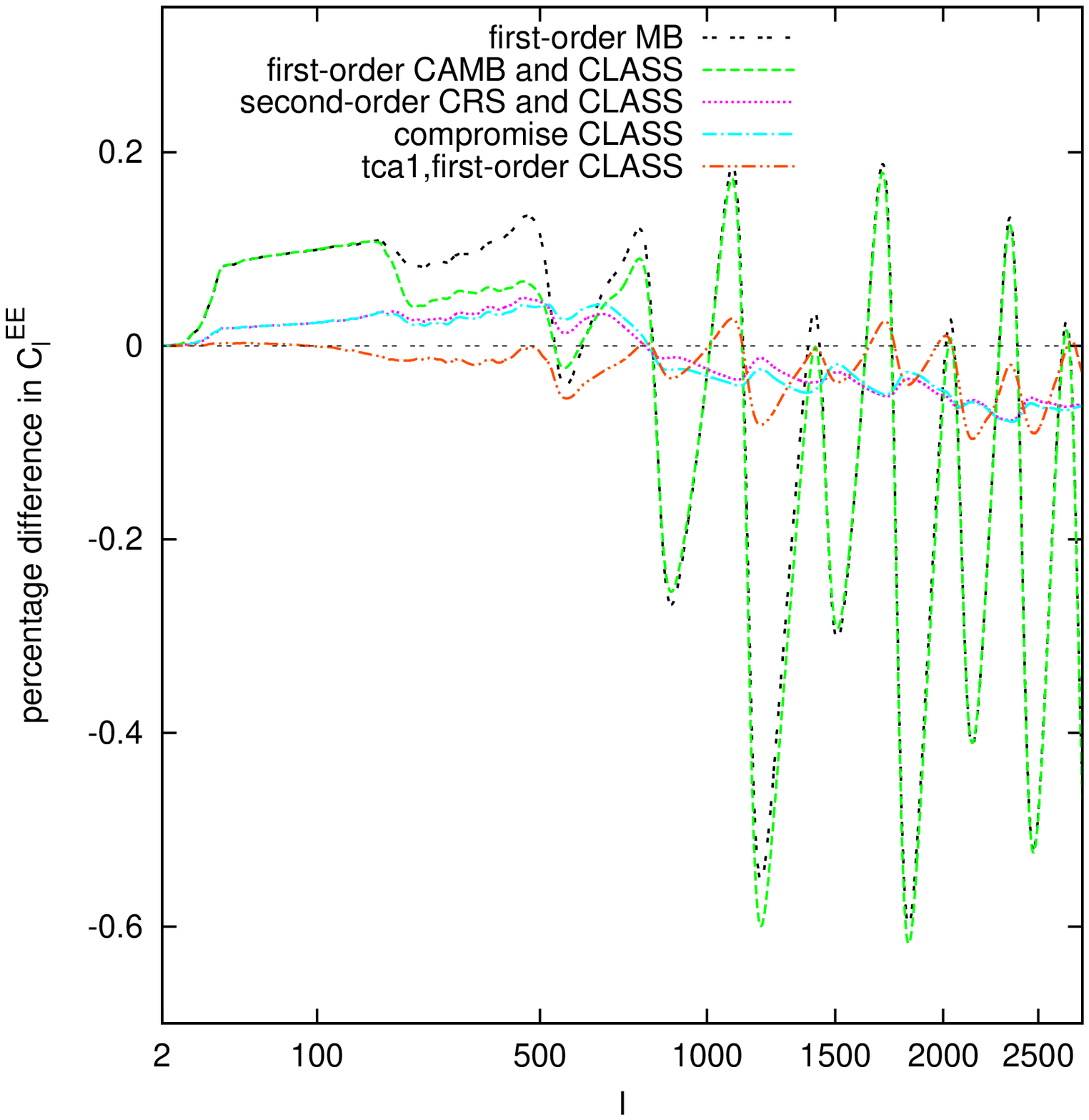}
\caption{\label{tca_cl2} {Impact of TCA schemes on temperature (left)
    and E-polarisation (right)}. For the various TCA schemes discussed
  in Sec.~\ref{tca_comp}, we show the power spectrum with accuracy
  settings {\tt tca3} compared to the reference spectrum {\tt no-tca}
  (see Table \ref{tca_tab} for precision parameter values). We show a
  single curve for cases which are indistinguishable by eye, namely:
  {\tt first\_order\_CAMB} and {\tt first\_order\_CLASS}; and also,
  {\tt second\_order\_CRS} and {\tt second\_order\_CLASS}. The default
  scheme {\tt compromise\_CLASS} is also hardly distinguishable from
  the {\tt second\_order\_CLASS}. The faint line shows for comparison
  the first-order results with accuracy {\tt tca1}: the error is then
  comparable to {\tt compromise\_CLASS} with {\tt tca3}.}
%\end{figure}
}

\TABLE{
%\begin{table}
\begin{tabular}{|l|c|c|c|c|}
\hline
precision & {\tt no-tca} & {\tt tca1} & {\tt tca3}\\
TCA scheme & (irrelevant) & {\tt first\_order\_class} & {\tt compromise\_class} \\
\hline
{\tt rk} & 1069s & 19.4s & 17.8s \\
{\tt ndf15} & 16s & 14.9s & 14.6s \\
\hline
\end{tabular}
\caption{\label{tca_time}
  Execution time of the perturbation module, in seconds, with several TCA settings and with the two integrators (Runge-Kutta and stiff integrator {\tt ndf15}). The last two columns lead to roughly the same level of accuracy.
 }
%\end{table}
}

%%%%%%%%%%%%%%%%%%%%%%%%%%%%%%%%%%%%
\section{Ultra-relativistic Fluid Approximation (UFA) \label{sec_ufa}}
%%%%%%%%%%%%%%%%%%%%%%%%%%%%%%%%%%%%

All massless neutrinos and ultra-relativistic relics can be treated as
a single species, labeled as ``\nur{}'' in \CLASS{}. The code assumes
that these species are fully decoupled. Hence, they just free-stream
within a given gravitational potential, and can be followed with the
collisionless Boltzmann equation expanded in harmonic space and
integrated over momentum \cite{Ma:1995ey}. The solution can be
formally written as the sum of spherical Bessel functions $j_l(k
\tau)$ (exhibiting damped oscillations for $\tau > l/k$), plus a
particular solution of the inhomogeneous equations sourced by metric
fluctuations. The fact that one part of the solution has an analytical
expression cannot be directly implemented in the code, because the
total perturbations $\delta_{\nur{}}$, $\theta_{\nur{}}$ and
$\sigma_{\nur{}}$ back-react on the metric perturbations through
Einstein equations, and affect the source terms in the Boltzmann
equation. We will use this decomposition only as a guideline for
deriving accurate approximation schemes.

%%%%%%%%%%%%%%%%%%%%%%%%%%%%%%%%%%%%
\subsection{Truncation of the Boltzmann hierarchy}
%%%%%%%%%%%%%%%%%%%%%%%%%%%%%%%%%%%%

Since the \nur{} species couple only gravitationally to other species,
we are only interested in tracking $\delta_{\nur{}}$, $\theta_{\nur{}}$
and $\sigma_{\nur{}}$. Higher multipoles must still be included since
they couple with the lower ones, but in all efficient Boltzmann codes,
the hierarchy is truncated at some low multipole value $l_{\rm
max}$. \CMBFAST{}, \CAMB{}, {\tt CMBEASY} and \CLASS{} all
use the truncation scheme proposed in Ma \& Bertschinger (Eq.~(51) of \cite{Ma:1995ey})
which is designed to minimize artificial reflection of power from
$l_{\rm max}$ back to lower multipoles.  Still, this truncation is
not perfect, and a significant amount of unphysical reflection cannot
be avoided for times beyond $\tau=l_{\rm max}/k$. This implies that in
order to compute an accurate CMB spectrum, $l_{\rm max}$ should be at
least of the order of 30.  The computation of the matter power spectrum
$P(k)$ on small scales, up to some wavenumber $k_{\rm max}$, is more
problematic: one should further increase $l_{\rm max}$ proportionally to
$k_{\rm max}$ in order to get converging results.

%%%%%%%%%%%%%%%%%%%%%%%%%%%%%%%%%%%%%%%%%%%%%%%%
\subsection{Sub-Hubble fluid approximation}
%%%%%%%%%%%%%%%%%%%%%%%%%%%%%%%%%%%%%%%%%%%%%%%%

The Ultra-relativistic Fluid Approximation (UFA) implemented in
\CLASS{} is based on the idea that for a given wavenumber, $l_{\max}$
should not necessarily be fixed throughout the whole time
evolution. The code considers two regimes: wavelengths larger or
comparable to the Hubble radius, and wavelengths much smaller than the
Hubble radius. The transition between the two regime occurs for each
$k$ when the product $k \tau$ (equal to $\tau / \tau_k$ and coinciding
with $\tau_H/\tau_k$ during radiation domination) reaches
some threshold value that we call here $(k \tau)_{\rm ufa}$.
Typically, $(k \tau)_{\rm ufa}$ is chosen in the range from $10$ to
$50$, depending on the required precision. The full name of this
parameter in the code is {\tt
  ur\_fluid\_trigger\_tau\_over\_tau\_k}. In the first regime $k \tau
\leq (k \tau)_{\rm ufa}$, the Boltzmann hierarchy can be truncated at some
$l_{\max}$ which can be chosen to be rather small: it is enough to
take to $l_{\max} \sim (k \tau)_{\rm ufa}$, since multipoles with $l > k
\tau$ are negligible (according to the spherical Bessel
function approximation). In the second regime $k \tau \geq (k \tau)_{\rm
  ufa}$, the code still follows the three variables $\delta_{\nur{}}$,
$\theta_{\nur{}}$ and $\sigma_{\nur{}}$, which are sourced by metric
perturbations. But multipoles in the range $2 < l \ll k \tau$ are
suppressed, leading to an effective decoupling between the first three
multipoles and the highest ones. Hence it is natural to lower $l_{\rm
  max}$ down to two in this regime. Ultra-relativistic neutrinos are
then described by a reduced system of equations for $\delta_{\nur{}}$,
$\theta_{\nur{}}$ and $\sigma_{\nur{}}$, i.e. by fluid equations (of
course, this fluid is not assumed to be perfect, since it has
anisotropic pressure). In summary, the UFA approximation consists in
lowering $l_{\max}$ from a value close to $(k \tau)_{\rm ufa}$ down to
$l_{\max}=2$ deep inside the Hubble radius, at the time when
$k \tau=(k \tau)_{\rm ufa}$. Such a scheme offers many advantages:
\begin{enumerate}
\item
  When computing the matter power spectrum, the number of \nur{}
  equations to integrate before the approximation is switched on does
  not need to be scaled linearly with the highest wave number $k_{\rm
  max}$.
\item The number of \nur{} equations reduces to $(l_{\max}+1)=3$ in the whole region of
  $(k, \tau)$ space fulfilling the condition $k \tau > (k \tau)_{\rm ufa}$;
  this is precisely the region in which the computation would be
  time-consuming, since {\nur{}} perturbations oscillate inside the
  Hubble radius.
\item The UFA completely avoids the issue of power reflecting at some
large $l_{\max}$, which would otherwise affect the evolution of low
multipoles periodically due to some spurious wave travelling back and forth
between $l=l_{\max}$ and $l=0$. (This behaviour is clearly seen in Fig.~\ref{nfa_var}).
\end{enumerate}  
The fluid approximation could in principle be used until present time,
but the code allows a more aggressive approximation, the Radiation
Streaming Approximation, to take over from the UFA after photon
decoupling. This new approximation is discussed in the next
section. Hence, the UFA is essentially a way to save computing time
during radiation domination and at the beginning of matter domination.

%%%%%%%%%%%%%%%%%%%%%%%%%%%%%%%%%%%%
\subsection{Fluid equations}
%%%%%%%%%%%%%%%%%%%%%%%%%%%%%%%%%%%%

We need to find a closed system for the evolution of $\delta_{\nur{}}$,
$\theta_{\nur{}}$ and $\sigma_{\nur{}}$, valid deep inside the Hubble
radius. The full system of equations in the synchronous gauge (Eq. (49) in
\cite{Ma:1995ey}) reads:
\bseq
\label{eq:systemur}
\begin{eqnarray}
&\delta_\nur{}'=-\frac{4}{3}\theta_\nur{}-\frac{2}{3}h', \label{deltanurdot}\\
&\theta_\nur{}'=k^2\left(\frac{1}{4}\delta_{\nur{}}-\sigma_{\nur{}}\right),\label{thetanurdot}\\
&2\sigma_\nur{}'=\frac{8}{15}\theta_\nur-\frac{3}{5} k F_{\nur 3}+\frac{4}{15}(h'+6\eta'),\label{shearnurdot}\\
&F_{\nur{}~l}'=\frac{k}{2l+1}\left[l F_{\nur{}( l-1)}-(l+1)F_{\nur{} (l+1)}\right].
\end{eqnarray}
\eseq
In Appendix~\ref{appendix:fluid}, we use the formal solution of these equations in order to derive an exact integral relation between $\sigma_\nur{}'$, $\sigma_\nur{}$, $\theta_\nur{}$ and metric perturbations. We then find an approximate but more practical form of this relation valid 
inside the Hubble radius, at leading order in an expansion in metric perturbation derivatives 
($h^{(n)}/k^{n-1}$, $\eta^{(n)}/k^{n-1}$) and in powers of $(k\tau)^{-1}$:
\begin{equation}
{\sigma_\nur{}}' = -\frac{3}{\tau}\sigma_\nur{}+\frac{2}{3}\theta_\nur{}
+\frac{1}{3}h'~.\label{ufa_class}
\end{equation}
Since metric perturbation only evolve over a Hubble time scale inside the Hubble radius, we expect this 
expansion to converge, and we will see below that the above
relation is indeed accurate enough for our purpose.
In the default version of \CLASS{}, this equation is used for closing the
system of equations when the UFA is switched on. This method corresponds to the
setting {\tt ufa\_method = ufa\_class} in the code's precision
parameter structure.

%%%%%%%%%%%%%%%%%%%%%%%%%%%%%%%%%%%%
\subsection{Alternative schemes}
%%%%%%%%%%%%%%%%%%%%%%%%%%%%%%%%%%%%

Some nearly equivalent schemes can be justified in slightly different ways. 
Truncating the Boltzmann equations at $l_{\rm max}=2$ with the usual
truncation scheme of Ma \& Bertschinger gives:
\begin{equation}
{\sigma_\nur{}}' = -\frac{3}{\tau}\sigma_\nur{}+\frac{2}{3}\theta_\nur{}
+\frac{1}{3}(h'+6\eta')~.\label{ufa_mb}
\end{equation}
This truncation scheme is based entirely on the assumption that $F_{\nur{}~l}(k,\tau) \propto j_l(k\tau)$.
So, the reason for the difference between eq.~(\ref{ufa_class}) and (\ref{ufa_mb}) is that
(\ref{ufa_class}) is based on the full formal solution, including the leading order contribution to the part sourced by the metric, while (\ref{ufa_mb}) is based only on the solution of the homogeneous equation.
Equation (\ref{ufa_mb}) is used when the user switches to {\tt ufa\_method =
  ufa\_mb}, and amounts to adding an extra term in $\eta'$. 
  %Although
  %one has $\eta' \ll h'$ inside the Hubble radius, using this method
  %instead of the default one gives a small difference in the final
  %result, because the term $\frac{1}{3}h'$ in (\ref{ufa_class}) and (\ref{ufa_mb}) almost
   %cancels with the term $\frac{2}{3}\theta_\nur{}$
  %(this can be checked in the previous analytic solution). 

  Finally, in a more general context, Hu \cite{Hu:1998kj} introduced a
  set of equations modeling a cosmological viscous fluid, and
  suggested that this fluid could approximate the evolution of
  free-streaming neutrinos with the parameter choice $(w, c_s^2,
  c_{\rm vis}^2) = (1/3, 1/3, 1/3)$. In this limit, Hu's fluid
  equations are identical to our UFA equations except for the shear derivative:
\begin{equation}
{\sigma_\nur{}}' = -3\frac{a'}{a}\sigma_\nur{}+\frac{2}{3}\theta_\nur{}
+\frac{1}{3}(h'+6\eta')~.\label{ufa_hu}
\end{equation}
The coefficient $-3\frac{a'}{a}$ reduces to $-3/\tau$ deep inside the
radiation dominated regime, but becomes different around the time of
equality.  This equation is used when the user switches to {\tt
ufa\_method = ufa\_hu}. Below we will compare the performances of
equations (\ref{ufa_class}, \ref{ufa_mb}, \ref{ufa_hu}) and show that
the first one is slightly more precise (as expected from the 
rigorous mathematical proof of Appendix~\ref{appendix:fluid}).

Finally, when the user selects {\tt ufa\_method = ufa\_none}, no UFA
scheme is employed, and the truncation multipole $l_{\rm max}$ remains
the same throughout the evolution.

%%%%%%%%%%%%%%%%%%%%%%%%%%%%%%%%%%%%
\subsection{Comparison at the level of perturbations}
%%%%%%%%%%%%%%%%%%%%%%%%%%%%%%%%%%%%

In Fig.~\ref{nfa_var}, we compare the evolution of $\delta_{\nur{}}$
and $\sigma_{\nur{}}$ for a given mode, obtained either by solving the
full Boltzmann equation up to a very high $l_{\rm max} \sim 3000$, or
with $l_{\rm max}=46$ with/without the default UFA. In absence of
approximation, one can see some spurious evolution appearing
periodically (here, around $k \tau \sim 100$ and $k \tau \sim 200$):
this corresponds to the propagation of an unphysical wave between the
multipole boundaries $l=l_\mathrm{max}$ and $l=0$. Using the default
UFA scheme {\tt ufa\_class}, we reproduce accurately the phase, the
amplitude and, to a lesser extent, the zero-point of the oscillations.
For the clarity of the figure, we do not show the results from
alternative approximation schemes. We checked that the {\tt ufa\_mb}
scheme also reproduces the correct phase and amplitude, but introduces a larger
error in the zero-point of oscillations. Finally, the {\tt ufa\_hu}
scheme reproduces the phase, but not the correct amplitude of the
oscillations.

\FIGURE{
%\begin{figure}
%
\includegraphics[width=\wi]{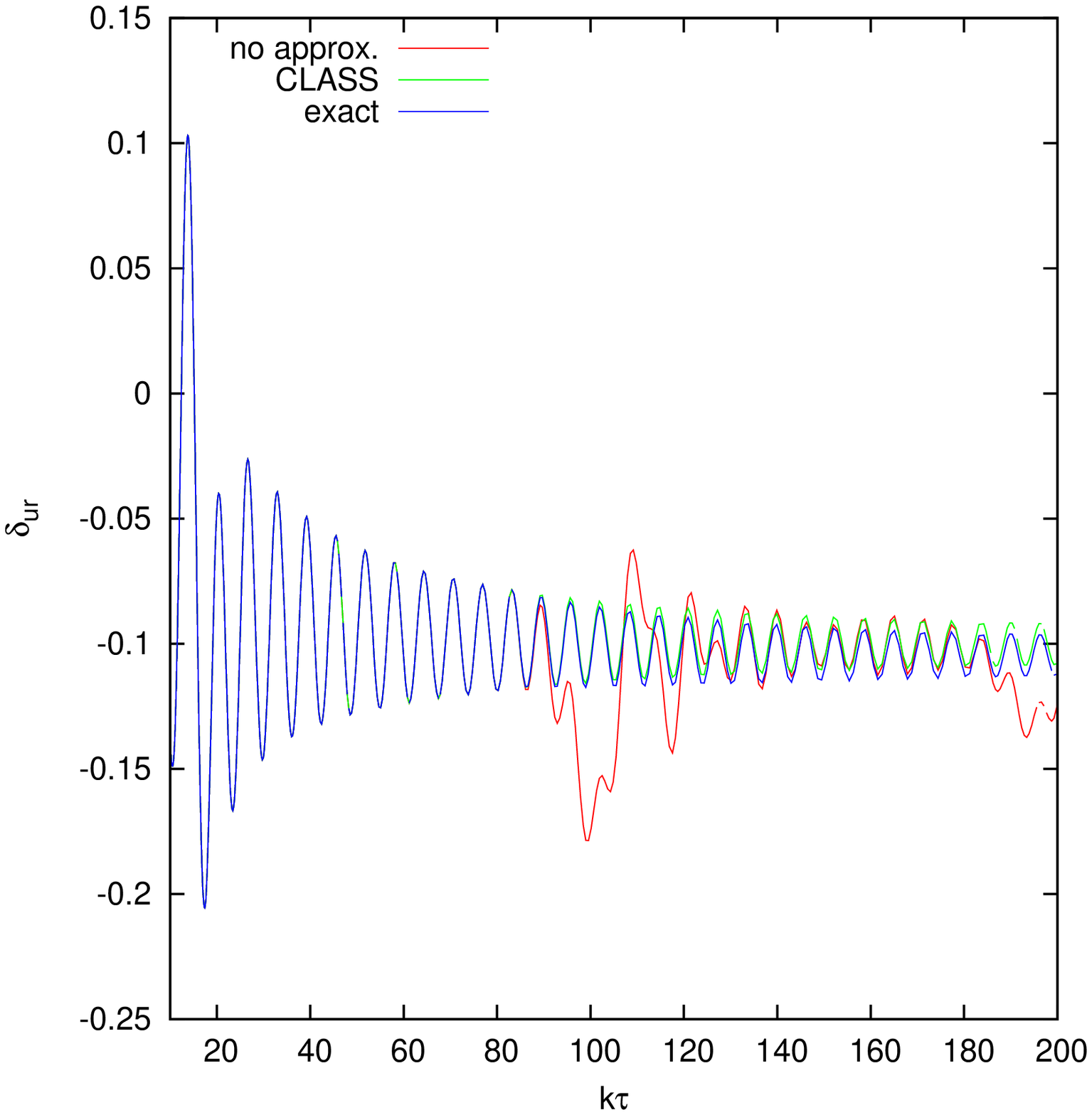}
\includegraphics[width=\wi]{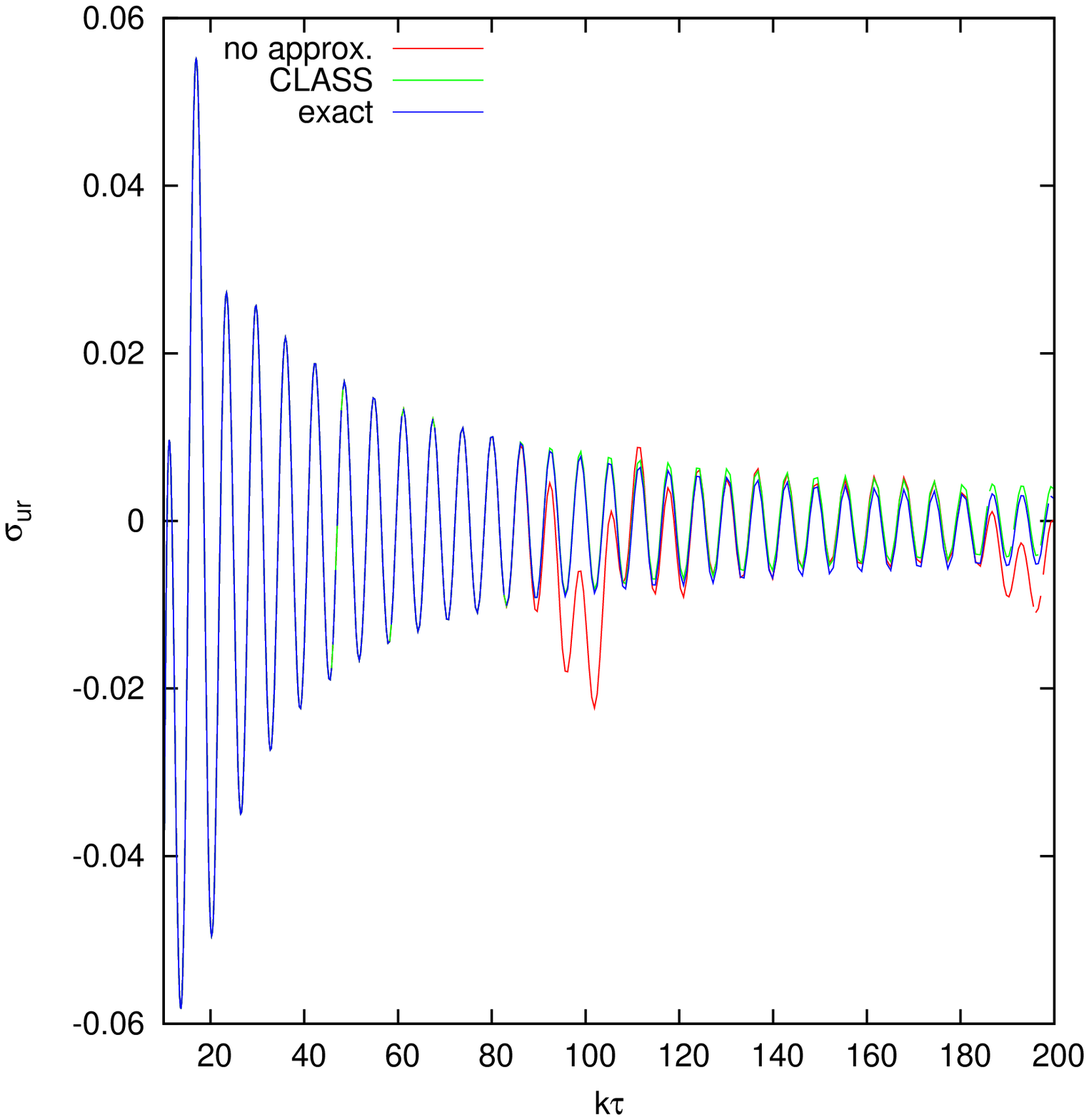}
\caption{\label{nfa_var} Evolution of ${\delta}_{\nur{}}$ (left) and
  $\sigma_{\nur{}}$ (right) for the mode $k=10^{-1}h{\rm Mpc}^{-1}$ over
  the range $10 < k \tau < 200$, i.e. well inside the Hubble radius
  and throughout the radiation dominated stage. In the ``exact'' case,
  the full Boltzmann equation for \nur{} is truncated at $l_{\rm max}
  \sim 3000$, with no impact of the truncation on the result. In the
  ``no approx.'' case, the truncation is performed at $l_{\rm
  max}=46$. In the ``\CLASS{}'' case, we use the UFA scheme {\tt ufa\_class} and set $l_{\rm max}=46$ as long as $k \tau \leq (k \tau)_{\rm ufa}=50$, 
  or $l_{\rm max}=2$ afterward.}
%\end{figure}
}

%%%%%%%%%%%%%%%%%%%%%%%%%%%%%%%%%%%%
\subsection{Comparison at the level of CMB/matter power spectrum}
%%%%%%%%%%%%%%%%%%%%%%%%%%%%%%%%%%%%

\FIGURE{
%\begin{figure}
%
\includegraphics[width=\wi]{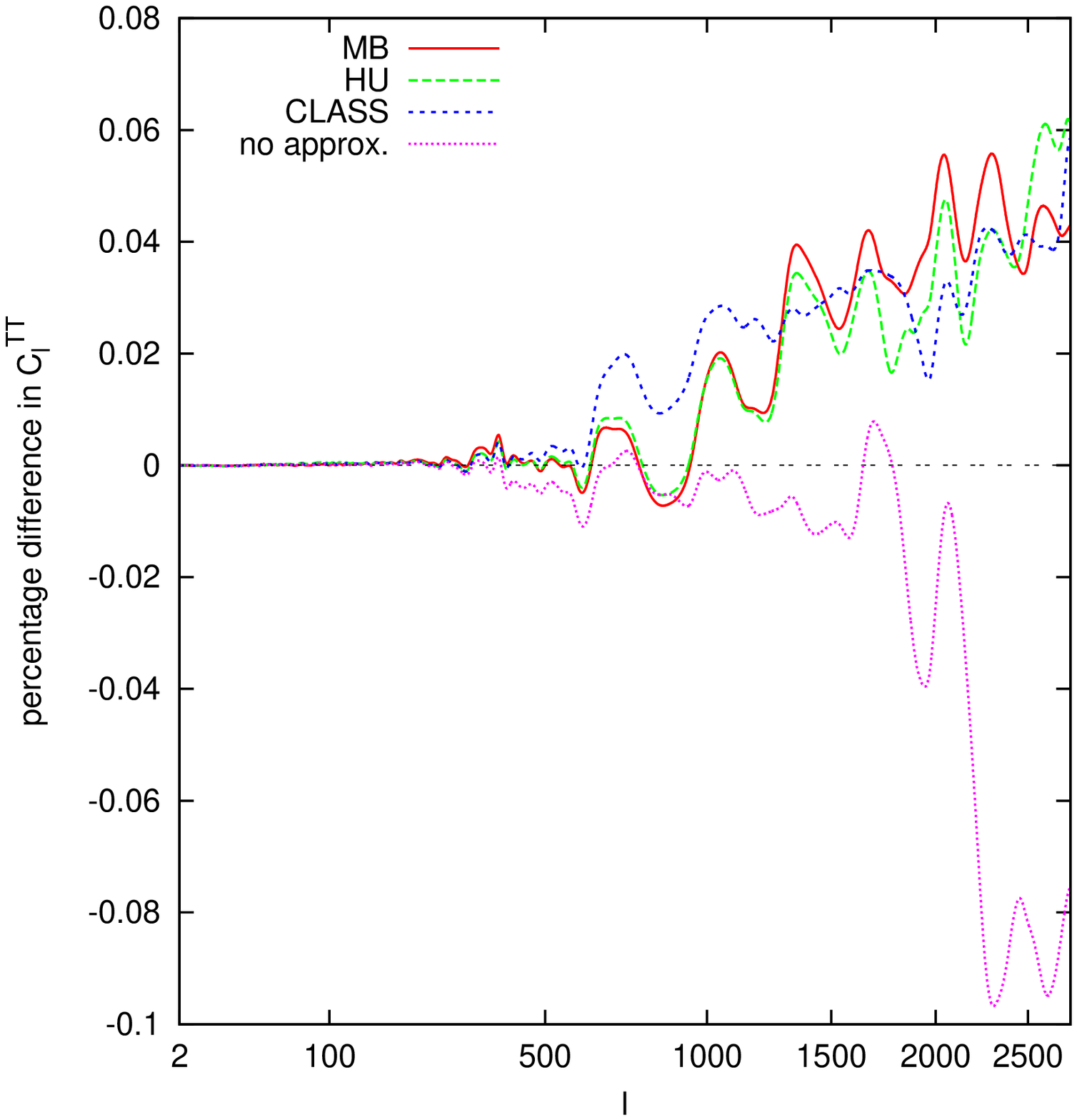}
\includegraphics[width=\wi]{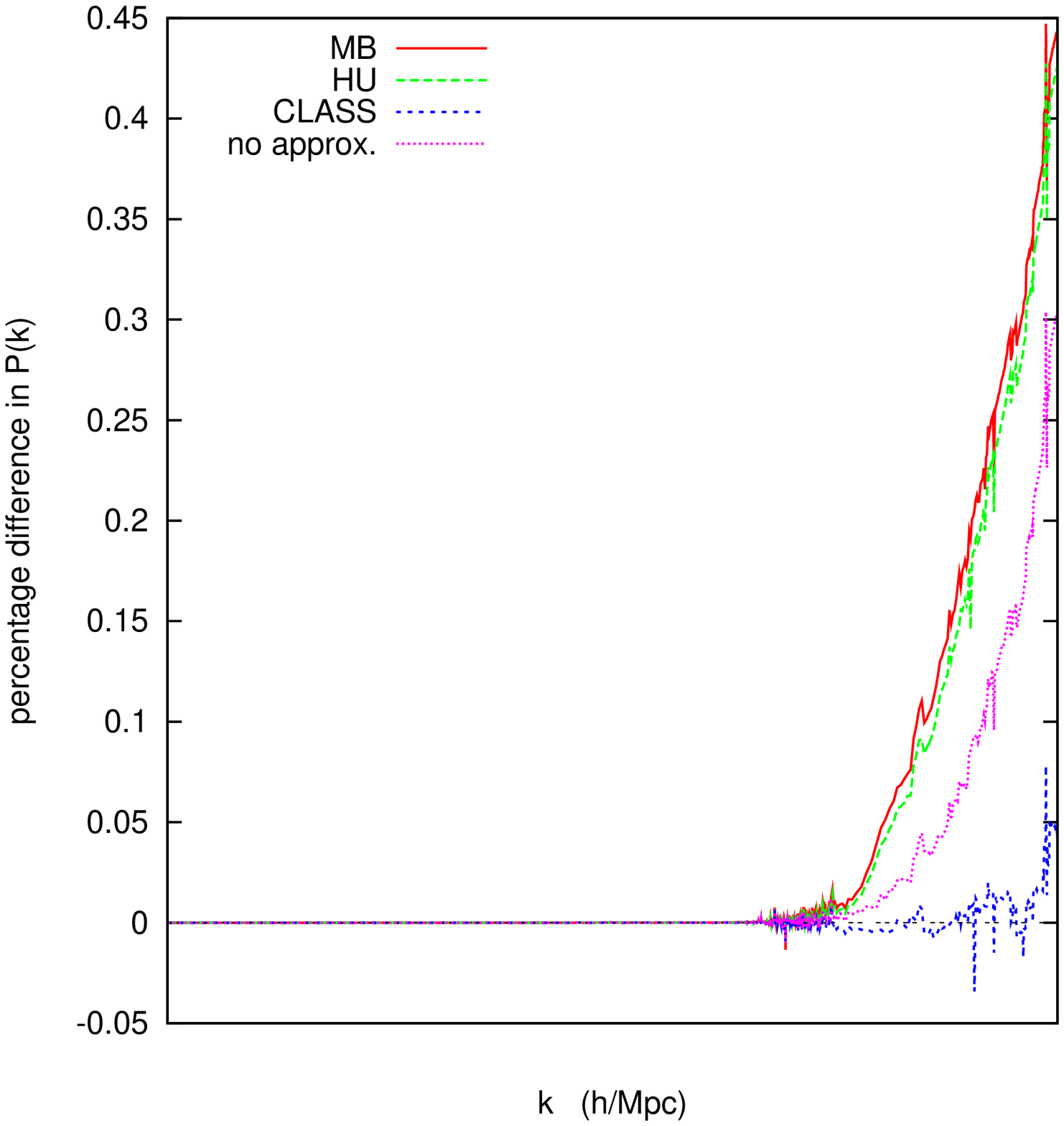}
\caption{\label{ufa_spec} Power spectra (for temperature CMB
anisotropies and for matter) obtained with various implementations of
the UFA, compared to those obtained in a reference run.
In the four cases displayed
here, we use $l_{\rm max}=18$ as long as $k \tau \leq (k \tau)_{\rm
ufa}=18$.  The curves labeled ``MB'', ``HU'', ``\CLASS{}'' correspond to
the three possible implementation of the UFA discussed in the text;
the last curve uses no UFA and a standard truncation scheme at $l_{\rm
max}=18$ until the time at which the next approximation RSA takes over
(see Sec.~\ref{sec_rsa}): in this case the code is at the same time
a bit less precise and 10\% slower. We do not show the results
for the polarisation spectrum $C_l^{EE}$, that look very similar to
those for $C_l^{TT}$.}
%\end{figure}
}

\TABLE{
%\begin{table}
\begin{tabular}{|l|c|c|}
\hline
UFA scheme (fixed $l_{\rm max}$ and $(k \tau)_{\rm ufa}$) & {\tt ufa\_none} & {\tt ufa\_class} \\
\hline
{\tt rk} & 29.7s & 27.0s  \\
{\tt ndf15} & 16.7s & 15.2s \\
\hline
\end{tabular}
\caption{\label{ufa_time}
  Execution time of the perturbation module in seconds, with the
  precision parameters of the file {\tt cl\_permille.pre}, plus
  $l_{\rm max}=(k \tau)_{\rm ufa}=18$. Using an ultra-relativistic
  fluid approximation leads simultaneously to a 10\% faster execution
  and to more accurate results.}
%\end{table}
}

In Fig.~\ref{ufa_spec}, we show the impact of the UFA on the CMB and
matter power spectrum. We take the precision parameters of the file
{\tt cl\_permille.pre}, and play with the values of $l_{\rm max}$ and
$(k \tau)_{\rm ufa}$, called {\tt l\_max\_ur} and {\tt
ur\_fluid\_trigger\_tau\_over\_tau\_k} in the code.  We first compute
some reference spectra with $l_{\rm max}=3000$ (to remove any
truncation effect), and such a large value of $(k \tau)_{\rm ufa}$
that the UFA is never used.  All other results from this section are
compared to these spectra. We then fix both $l_{\rm max}$ and $(k
\tau)_{\rm ufa}$ to 18 and vary only the {\tt
ur\_fluid\_approximation} parameter. We show the error induced by each
UFA scheme for the temperature and matter power spectrum in
Fig.~\ref{ufa_spec} (results for temperature and polarisation are very
similar).  The results from the {\tt ufa\_none} case are very unstable
and depend a lot on the choice of $l_{\rm max}$ and $(k \tau)_{\rm
ufa}$) values.  With the present choice, they correspond to a twice
larger error in the CMB spectra than in any UFA scheme; for slightly
different choices they would also induce a larger error in the matter
power spectrum. The three UFA schemes, which do not have such
instabilities, are nearly as good as each other for CMB spectra, while
for the matter power spectrum the {\tt ufa\_class} scheme is one order
of magnitude better. Table \ref{ufa_time} shows that the UFA
approximation allows for a 10\% speed up, while being more accurate
for a fixed $l_{\rm max}$.

%%%%%%%%%%%%%%%%%%%%%%%%%%%%%%%%%%%%
\section{Radiation Streaming Approximation (RSA)\label{sec_rsa}}
%%%%%%%%%%%%%%%%%%%%%%%%%%%%%%%%%%%%

After their respective decoupling time, the photons gradually
free-stream like neutrinos (except around the time of reionisation at
which their coupling to baryons is enhanced). In principle, it would
be possible to look for a fluid approximation for photons, like we did
for neutrinos in the previous section. However we can go further than
that, since during this period the universe is dominated by matter and
eventually $\Lambda$/Dark Energy: in this case photons and massless
neutrinos almost behave like test-particles in an external
gravitational field, and we do not need to catch their evolution with
high accuracy (which was not the case for the \nur{} species during
radiation domination).

Like for massless neutrinos, in all efficient Boltzmann codes, the
Boltzmann equation for photons is truncated at some low multipole
value $l_{\rm max}$ using the truncation scheme proposed in Ma \&
Bertschinger (Eqs.~(65) in \cite{Ma:1995ey}). If $l_{\rm max}$ is not large enough, the
spurious reflection of power induced by the truncation propagates to
the final results, because radiation perturbations still play a small 
role during the free-streaming regime. More precisely:
\begin{itemize}
\item
the photon density fluctuation, velocity and shear perturbations appear
in Einstein equations;
\item
the photon density fluctuation and shear appear in the temperature/polarisation
source functions;
\item 
the photon velocity appears in the evolution equations of baryons
(since the baryon-photon coupling is not negligible during
reionisation).
\end{itemize}
In order to avoid propagating such an error, one can either increase
$l_{\rm max}$, or find a way to infer the photon density,
velocity and shear from some analytic Radiation Streaming
Approximation (RSA), in which case the integration of Boltzmann equations
can be stopped soon after photon decoupling.
Mathematically, this analytic
approximation should coincide with the particular non-oscillatory
solution of the inhomogeneous Boltzmann equations. Once the damped
oscillations accounted for by spherical Bessel functions becomes negligible
(i.e. when $k \tau \gg l$), the analytic approximation will coincide
with the true solution. Before that time, it will provide a correct
approximation to the true quantities averaged over a few oscillations.
In summary, the RSA has a double goal: to
avoid unphysical oscillations created by the Boltzmann truncation, and
to avoid wasting time in integrating the Boltzmann equations over many
such oscillations.

The RSA does not need to be accurate at very late time (end of matter
domination, $\Lambda$/Dark Energy domination), since by then radiation
fluctuations are always completely negligible with respect to matter
fluctuations. However, it should be reasonably accurate soon after
photon decoupling, i.e. during matter domination, when the energy
density in radiation, $\Omega_r\equiv \Omega_\gamma +\Omega_{\nur{}}$,
is smaller than one but not yet much smaller. The advantage of a
better approximation is two-fold. First, it can be switched on
earlier. Second, before switching on the approximation, we can use a
smaller value of $l_\text{max}$, since high multipoles will not have
time to grow.

The same treatment can be applied to ultra-relativistic species, which
are identical to photons in this regime, except for the fact that
they remain collisionless during reionisation. When the RSA is turned
on for photons, it is better to follow ultra-relativistic species in
the same way as photons, rather than with the fluid formalism
described in Sec.~\ref{sec_ufa}. The RSA then removes three more
differential equations, and cures the fact that the UFA turns out to
be inaccurate at late time. Hence, the default version of \CLASS{} treats
ultra-relativistic species first with exact equations, then with the
UFA (inside the Hubble radius and until photon decoupling), and
finally with the RSA (inside the Hubble radius and after photon
decoupling).

An expression for the RSA was discussed in the Newtonian gauge by
Doran \cite{Doran:2005ep}. Soon after, a somewhat simpler RSA
(neglecting reionisation) was also introduced in the \CAMB{} code, which
uses the synchronous gauge. Here, we will derive  an
approximation comparable to that of Doran \cite{Doran:2005ep}, but valid in the
synchronous gauge.

%%%%%%%%%%%%%%%%%%%%%%%%%%%%%%%%%%%%
\subsection{Relativistic relics (massless neutrinos)}
%%%%%%%%%%%%%%%%%%%%%%%%%%%%%%%%%%%%

We start with the simplest case, that of ultra-relativistic species
\nur{}. We combine the first two equations of (\ref{eq:systemur}) into
\begin{equation}
{\delta_{\nur{}}}'' + \frac{k^2}{3} \delta_{\nur{}}
= - \frac{2}{3} {h''} + \frac{4}{3} k^2 \sigma_{\nur{}}.
\end{equation} 
Inside the Hubble scale (i.e. when $k \tau \gg 1$) we can assume in
first approximation that $|\sigma_{\nur{}}| \ll |\delta_{\nur{}}|$ and neglect
the shear in the RSA. Also, since we are
looking for a smooth (non-oscillatory) particular solution of this
inhomogeneous equation, we can assume that $|{\delta''}_{\nur{}}| \ll k^2
|\delta_{\nur{}}|$. We conclude that the RSA for $\delta_{\nur{}}$ is simply
\begin{equation}
\delta_{\nur{}} = - \frac{2}{k^2} h''~. \label{fsa_delta_nur}
\end{equation}
Note that in the synchronous gauge, $h'$ coincides with $-2
\delta_{\cdm{}}'$, where $\delta_\cdm$ is the cold dark matter density contrast. Deep inside the matter-dominated regime, $\delta_{\cdm{}}
\propto a \propto \tau^2$, so $h'$ is linear in $\tau$, and $h''$ is a
constant. The RSA for $\delta_{\nur{}}$ is
therefore nearly static.
Concerning $\theta_{\nur{}}$, its value in the RSA is given by the exact
energy-conservation equation ${\delta'}_{\nur{}}=-\frac{4}{3}
\theta_{\nur{}}-\frac{2}{3} h'=0$, namely
\begin{equation}
\theta_{\nur{}}=-\frac{1}{2} h'~. \label{fsa_theta_nur}
\end{equation}
In practice, we must extract $h'$ and $h''$ from the Einstein equations
in the synchronous gauge, that read
\bseq
\begin{eqnarray}
2k^2 \eta - \frac{a'}{a} h' &=& 8 \pi {\cal G} a^2 \delta \rho_\text{tot}, \\
2 k^2 \eta' &=& 8 \pi {\cal G} a^2 [(\bar{\rho} + \bar{p}) \theta]_\text{tot}, \\
h'' + 2   \frac{a'}{a} h' - 2 k^2 \eta &=& - 8 \pi {\cal G} a^2 \delta p_\text{tot}~.
\end{eqnarray}
\eseq
Here we do not need the fourth equation sourced by the shear. The
difficulty comes from the fact that in order to infer $\delta_{\nur{}}$
we should compute $h''$, and for doing that we need to combine the
first and third equation, i.e. we need to know $ \delta \rho_\text{tot}$,
which depends itself on $\delta_{\nur{}}$. Fortunately, we can notice
that if we omit $\delta_{\nur{}}$ in the computation of $ \delta
\rho_\text{tot}$, we make a tiny error, since during matter domination
$|\delta \rho_{\nur{}}| \ll |\delta \rho_{\cdm{}}|$. Hence, it is good enough
to evaluate the first Einstein equation with $\delta_{\nur{}}$ set to
zero. The same is not true for the second equation, since the
synchronous gauge is comoving with \cdm{}, so one has $\theta_{\cdm{}}=0$ by
construction. As a result, neglecting $\theta_{\nur{}}$ in the
computation of $\theta_\text{tot}$ and $\eta'$ leads to a significant
inaccuracy in the solution for $\eta$.  Hence, we introduce the
following scheme:
\begin{enumerate}
\item
we compute $\delta \rho_\text{tot}$ assuming $\delta_{\nur{}}=0$, and obtain
$2k^2 \eta - \frac{a'}{a} h'$ from the first Einstein equation.
\item
using the fact that during matter domination $|\delta p_\text{tot}| \ll |\delta \rho_\text{tot}|$,
we notice that
\begin{equation}
\left| h'' + 2   \frac{a'}{a} h' - 2 k^2 \eta \right| \ll
\left| \frac{a'}{a} h' - 2k^2 \eta \right|,
\end{equation}
and hence to very good approximation
\begin{equation}
h'' = -2 \frac{a'}{a} h' + 2 k^2 \eta~. \label{fsa_hpp}
\end{equation}
We then infer the following RSA for $\delta_{\nur{}}$ from Eq.~(\ref{fsa_delta_nur}) :
\begin{equation}
\delta_{\nur{}} = \frac{4}{k^2} \left(\frac{a'}{a} h' - k^2 \eta \right)~. \label{fsa_delta_nur2}
\end{equation}
This formula is practical since $\eta$ is one of the variables that we integrate over time,
and $h'$ has been inferred in the previous step.
\item
We impose the free-streaming solution for $\theta_{\nur{}}$ (Eq.~(\ref{fsa_theta_nur})) and set $\sigma_{\nur{}}=0$.
\item
We use the remaining Einstein, continuity and Euler equations to evolve the system.
\end{enumerate}

%%%%%%%%%%%%%%%%%%%%%%%%%%%%%%%%%%%%
\subsection{Photons}
%%%%%%%%%%%%%%%%%%%%%%%%%%%%%%%%%%%%

For photons, the solution is a bit more complicated since the
baryon-photon coupling cannot be neglected during reionisation. 
We then need to find the particular non-oscillatory solution of (cf. \eqref{eq:photon})
\begin{equation}
{\delta''}_{\gamma} + \frac{k^2}{3} \delta_{\gamma}
= - \frac{2}{3} {h''} + \frac{4}{3} k^2 \sigma_{\gamma}
- \frac{4}{3\tau_c} (\theta_b-\theta_\gamma).
\end{equation}
Once again we will neglect the shear and search for a particular solution 
slowly varying with time ($|{\delta''}_{\gamma}| \ll |k^2 \delta_{\gamma}|$). In order to deal with
the coupling term, we expand the solution in powers of 
$\tau_c^{-1}$. The zeroth-order solution is exactly
similar to that of massless neutrinos, Eqs.~(\ref{fsa_delta_nur2}),
(\ref{fsa_theta_nur}). The first-order solution should satisfy
\begin{equation}
\frac{k^2}{3} \delta_{\gamma}
= - \frac{2}{3} {h''}
- \frac{4}{3\tau_c} (\theta_b+\frac{1}{2} h'),
\end{equation}
in which $h''$ can be replaced using Eq.~(\ref{fsa_hpp}). This approximation
turns out to work out very well, unlike the zeroth-order
solution. The velocity is then given by the exact energy-conservation equation
\begin{equation}
\theta_\gamma = -\frac{1}{2}h'-\frac{3}{4}{\delta'}_\gamma~.
\end{equation}
We take the derivative of the previous result for $\delta_{\gamma}$, assuming that
$h''$ is time-independent, and using once more
Eq.~(\ref{fsa_hpp}). We obtain:
\begin{equation}
\theta_\gamma =  -\frac{1}{2}h'
+\frac{3}{k^2\tau_c}\left[
-\frac{\tau_c'}{\tau_c}\left(\theta_b+\frac{1}{2} h'\right)
+  \left({\theta_b'}+\frac{1}{2} h''\right)
\right]~,
\end{equation}
in which $h''$ can be replaced using Eq.~(\ref{fsa_hpp}).
However the exact expression of $\theta_b'$ depends again on
$\theta_\gamma$. Like before, we use a perturbative scheme in
$\tau_c^{-1}$ and replace $\theta_b'$ above by its expression at first-order
in $\tau_c^{-1}$.

%%%%%%%%%%%%%%%%%%%%%%%%%%%%%%%%%%%%
\subsection{Summary of RSA equations}
%%%%%%%%%%%%%%%%%%%%%%%%%%%%%%%%%%%%

In summary, the RSA consists in neglecting
$\delta_{\gamma}$ and $\delta_{\nur{}}$ in the evolution of $\delta \rho_\text{tot}$ in the first
Einstein equation, which allows us to obtain $h'$, and then in imposing
\bseq
\begin{eqnarray}
\delta_{\gamma} &=&  
\frac{4}{k^2} \left(\frac{a'}{a} h' - k^2 \eta \right)
+ \frac{4}{k^2\tau_c} \left(\theta_b +\frac{1}{2} h' \right)~, \\
\theta_\gamma
&=& 
-\frac{1}{2}h'
+\frac{3}{k^2\tau_c}\left[
-\frac{\tau'_c}{\tau_c}\left(\theta_b+\frac{1}{2} h'\right)
+ \left(-\frac{a'}{a}\theta_b + c_b^2k^2 \delta_b
- \frac{a'}{a} h' + k^2 \eta\right)
\right],\\
\sigma_\gamma &=&0~,\\
\delta_{\nur{}}&=&
\frac{4}{k^2} \left(\frac{a'}{a} h' - k^2 \eta \right)~,\\
\theta_{\nur{}} &=& -\frac{1}{2}h'~, \\
\sigma_{\nur{}}&=&0~.
\end{eqnarray}
\eseq
This scheme is set to be the default one in \CLASS{}, as long as the
precision variable {\tt radiation\_streaming\_approximation} remains set to
{\tt rsa\_MD\_with\_reio}. For comparison, some cruder schemes can
be used: if the same variable is set to {\tt rsa\_MD}, the code
will use the above expressions at zero order in $\tau_c^{-1}$ (i.e, with
$\delta_\gamma=\delta_{\nur{}}$ and $\theta_\gamma=\theta_{\nur{}}$). If it
is set to {\tt rsa\_none}, the radiation perturbations are just set
to zero.

%%%%%%%%%%%%%%%%%%%%%%%%%%%%%%%%%%%%
\subsection{Comparison at the level of perturbations}
%%%%%%%%%%%%%%%%%%%%%%%%%%%%%%%%%%%%

In Fig.~\ref{fa_var}, we show the evolution of $\delta_\gamma$,
$\theta_\gamma$, $\theta_b$ and $\eta$ between recombination and
present time, for a particular wavenumber $k=0.1$~Mpc$^{-1}$. We
compare two RSA schemes with the exact evolution obtained by
integrating all multipoles at all times. We always keep the UFA
approximation of Sec.~\ref{sec_ufa} turned off.  We see that
immediately after switching on the RSA, our approximation for
$\delta_\gamma$ (and also for $\delta_\nur{}$, which is not shown)
matches accurately the exact evolution averaged over a few oscillations.
This would not be the case with several simpler RSA schemes which
assume full matter domination and an exact linear growth of
$h'(\tau)$.  In the models used for the figures, reionisation takes
place at $z_*=10$ and $\tau=4458$~Mpc. It induces a clear feature in
$\delta_\gamma$ and $\theta_b$ (having impact on $\eta$) which is well
captured by the terms proportional to $\tau_c^{-1}$ in the full {\tt
  rsa\_MD\_with\_reio} scheme.

\FIGURE{
%\begin{figure}
%
\includegraphics[width=\wi]{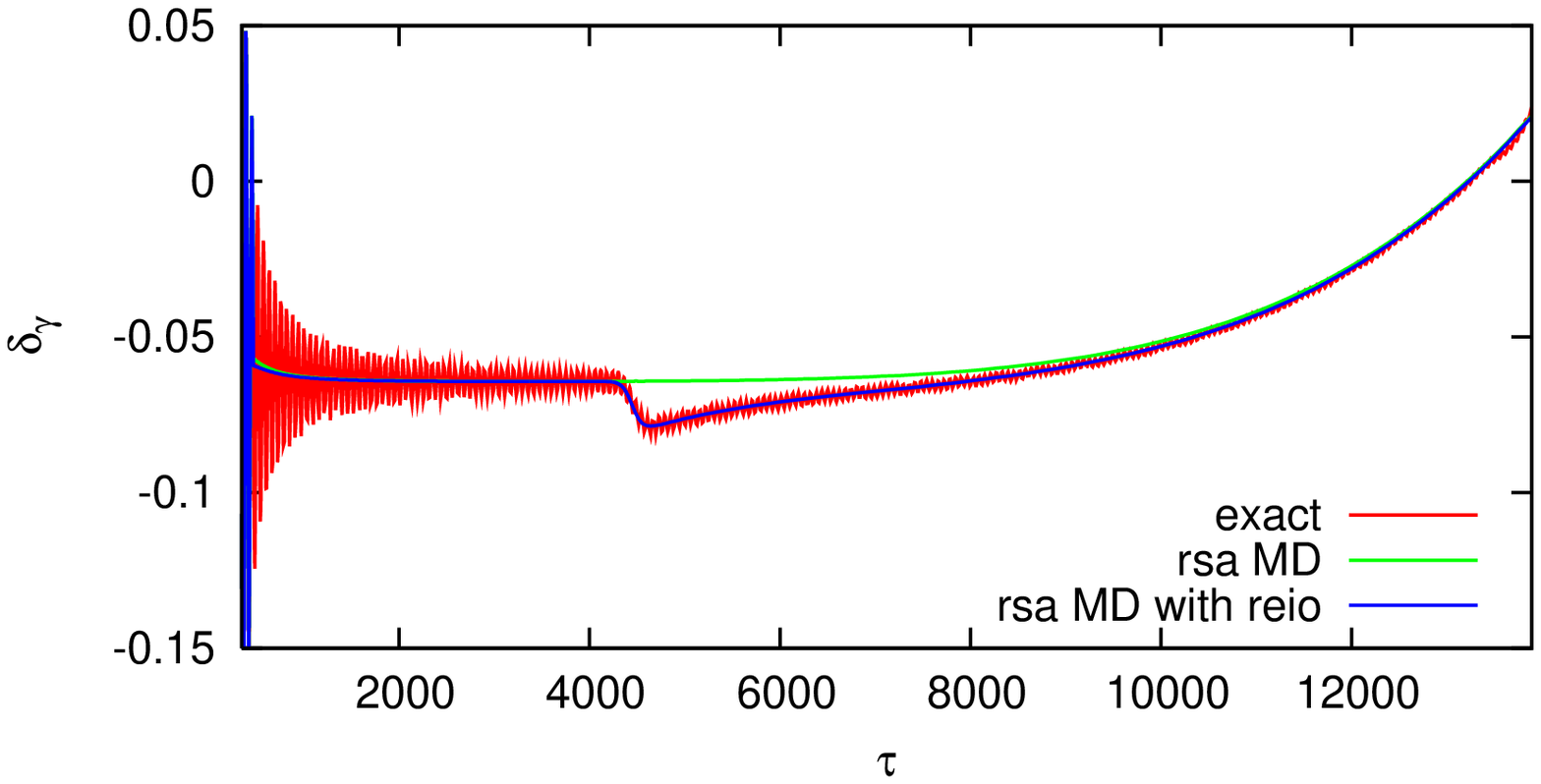}
\includegraphics[width=\wi]{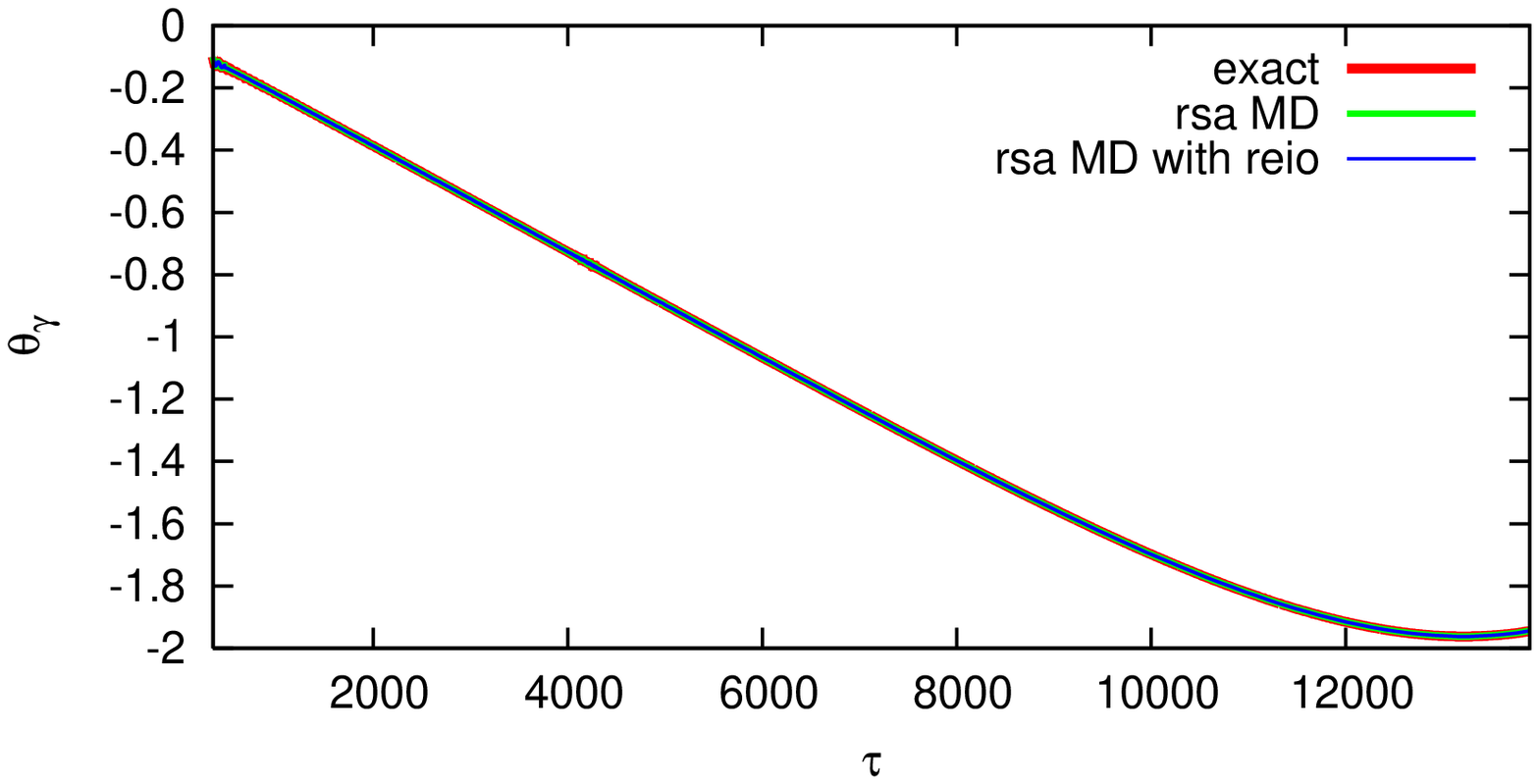}\\
\includegraphics[width=\wi]{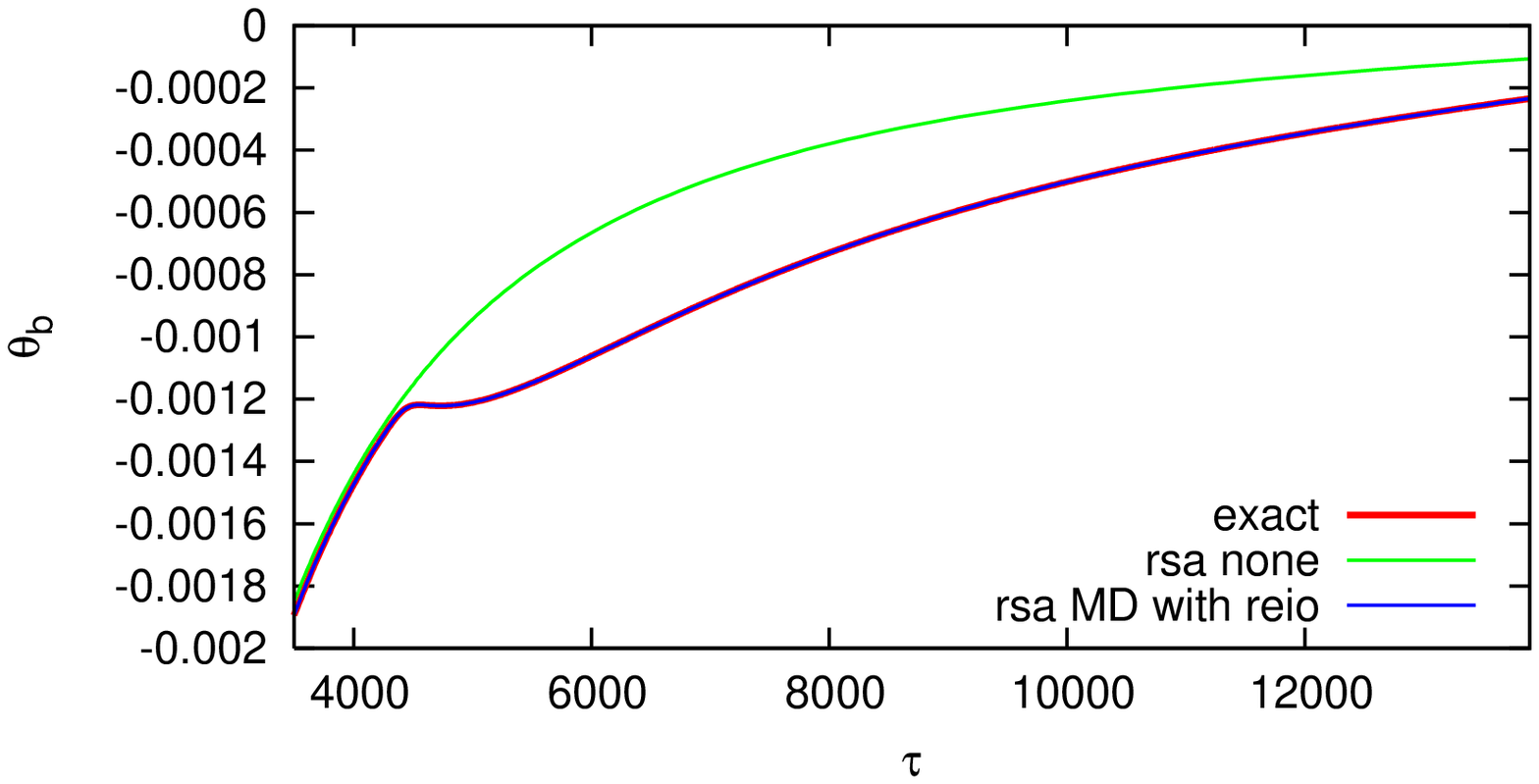}
\includegraphics[width=\wi]{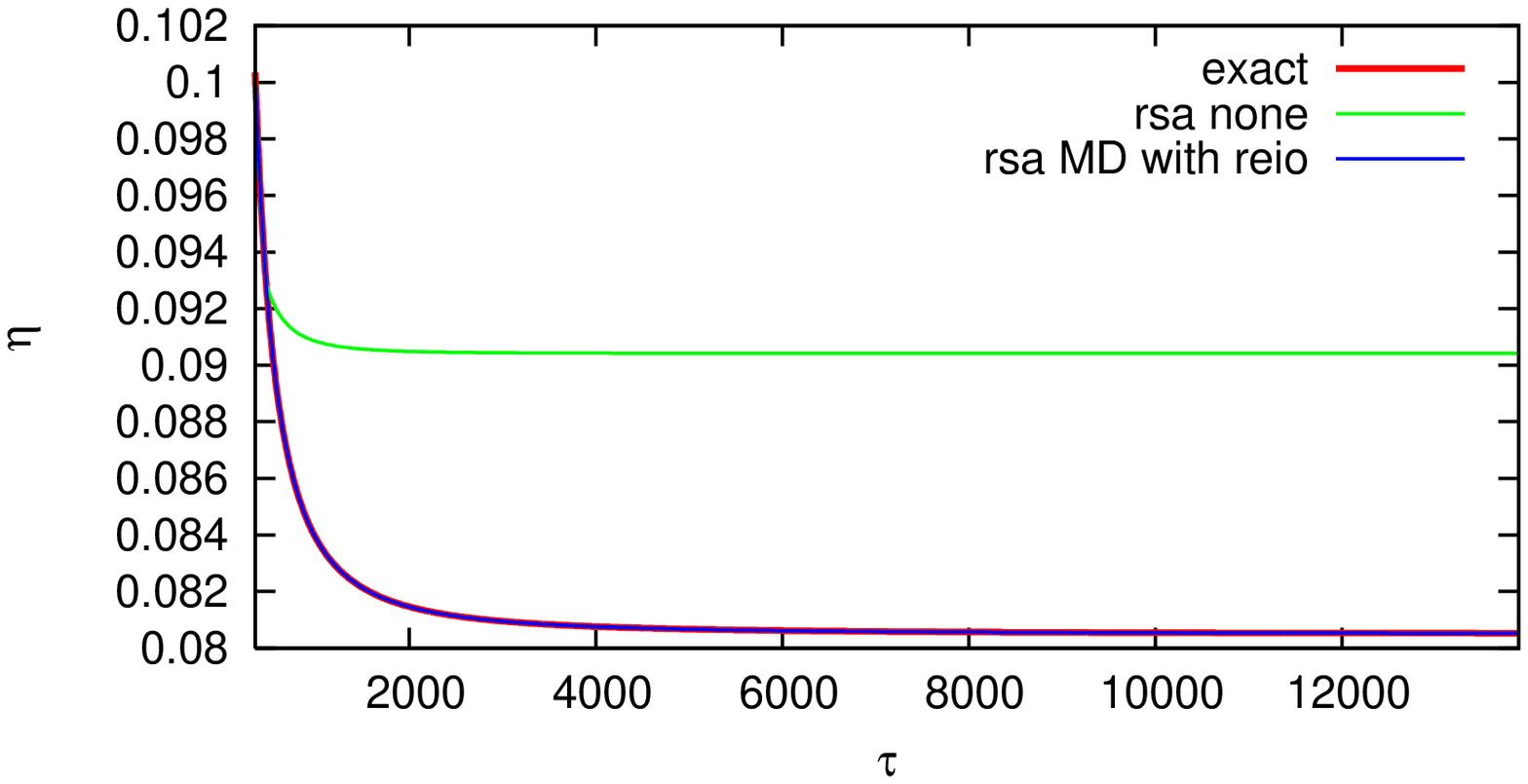}
\caption{\label{fa_var} Evolution of the quantities $\delta_\gamma$
  (top left), $\theta_\gamma$ (top right), $\theta_b$ (bottom left),
  $\eta$ (bottom right) for the mode $k=0.1$~Mpc$^{-1}$, as a function
  of conformal time (in Mpc), between a time chosen soon after photon
  decoupling (or slightly before reionisation in the $\theta_b$ plot)
  and today. The red curves show the result of an exact integration
  with no truncation or approximation. In the blue curves, the
  RSA is turned on around $\eta = 470$~Mpc,
  just before the solutions become unphysical due to the Boltzmann
  equation truncation at {\tt l\_max\_g}=12, {\tt l\_max\_pol\_g}=12,
  {\tt l\_max\_ur}=28. For comparison, in green, we show the result
  for $\delta_\gamma$ when the terms in $\tau_c^{-1}$ are neglected in the
  RSA, and those for $\theta_b$ and $\eta$
  when the radiation multipoles are all set to zero instead of using a free-streaming solution.  }
%\end{figure}
}

%%%%%%%%%%%%%%%%%%%%%%%%%%%%%%%%%%%%
\subsection{Comparison at the level of temperature/polarisation spectra}
%%%%%%%%%%%%%%%%%%%%%%%%%%%%%%%%%%%%

The precision parameters governing the evolution of perturbations in
the late universe are:
\begin{itemize}
\item two parameters defining the time at which the RSA is switched
  on. For each wavenumber, we stop evolving photon and \nur{}
  perturbations when the two conditions $$k \tau = \tau / \tau_k \geq
  {\tt radiation\_streaming\_trigger\_tau\_over\_tau\_k}$$ and
  $$\tau_c / \tau \geq {\tt
  radiation\_streaming\_trigger\_tau\_c\_over\_tau}$$ are satisfied
  (i.e., photons are sufficiently decoupled, and the wavelength is
  sufficiently deep inside the sub-Hubble regime). In principle,
  it would be possible to switch on the RSA at different times for
  photons and {\tt ur} species, but for simplicity we did not consider
  this option.
\item {\tt l\_max\_g}, {\tt l\_max\_pol\_g} and {\tt l\_max\_ur}
  define the number of photon temperature, photon polarisation and
  \nur{} multipoles which are integrated until the RSA is switched on.
\end{itemize}

In order to compute some reference spectra to be used throughout this
section, we fix the precision parameters according to the file {\tt
  cl\_permille.pre}, increase {\tt l\_max\_g}, {\tt l\_max\_pol\_g}
and {\tt l\_max\_ur} to 3000, and choose such large values of the
trigger parameters that the UFA and RSA are never employed.  We wish
to compare these reference spectra with those from runs with/without
the RSA, in which {\tt l\_max\_g}, {\tt l\_max\_pol\_g} and {\tt
  l\_max\_ur} are kept fixed to reasonable values.
We choose {\tt l\_max\_g} and {\tt l\_max\_pol\_g} to be equal to
18. Since we do not want to use the UFA approximation in this
comparison (in order to focus only on the impact of the RSA), we fix
{\tt l\_max\_ur} to a larger value, namely 50. This setting, called
{\tt no-rsa} in Table \ref{fa_tab}, leads to a 0.01\% error both in
the temperature multipoles and in the matter power spectrum for $k
\leq 1h$Mpc$^{-1}$, as shown in Fig.~\ref{fa_cl1}. Finally, in the
run called {\tt rsa}, we switch on the default RSA scheme, with the
trigger values specified in Table \ref{fa_tab}. With such settings,
the error in the temperature (and also polarisation) multipoles
remains as small as without the RSA, while the error in the matter
power spectrum grows moderately to 0.04\% (see
Fig.~\ref{fa_cl1}). However, the running time is reduced
considerably, as shown in Table \ref{fa_time}: with the Runge-Kutta
integrator, the RSA leads to a 66\% speed up. Note that the {\tt
  ndf15} remains much better than the Runge-Kutta integrator when the
RSA is employed (by a factor 2) while it experiences difficulties in
following oscillatory solutions in absence of a RSA. However, the
combination of our RSA scheme and {\tt ndf15} integrator leads to very
nice performances (speed-up by a factor 4 with respect to Runge-Kutta
without any RSA).

\TABLE{
%\begin{table}
\begin{tabular}{|l|c|c|c|}
\hline
precision setting: & reference & {\tt no-rsa} & {\tt rsa} \\
\hline
{\tt l\_max\_g} & 3000 & 18 & 18 \\
{\tt l\_max\_pol\_g}& 3000 & 18 & 18 \\
{\tt l\_max\_ur}& 3000 & 50 & 50 \\
{\tt radiation\_streaming\_trigger\_tau\_over\_tau\_k} & $\infty$ & $\infty$ & 100 \\
{\tt radiation\_streaming\_trigger\_tau\_c\_over\_tau} & $\infty$ & $\infty$ & 2 \\
{\tt ur\_fluid\_trigger\_tau\_over\_tau\_k} & $\infty$ & $\infty$ & $\infty$ \\
\hline
\end{tabular}
\caption{\label{fa_tab} Three settings for the parameters governing
  the Boltzmann truncation and RSA. All other parameters are fixed
  with the file {\tt cl\_permille.pre} of the public \CLASS{}
  distribution: in particular, {\tt
    radiation\_streaming\_approximation} is set to {\tt
    rsa\_MD\_with\_reio}. The reference run never uses the UFA and
  RSA, and cannot be affected by the Boltzmann truncation. The second
  and third settings share the same truncation multipoles, and differ only
  by using the RSA or not.}
%\end{table}
}

\FIGURE{
%\begin{figure}
%
\includegraphics[width=\wi]{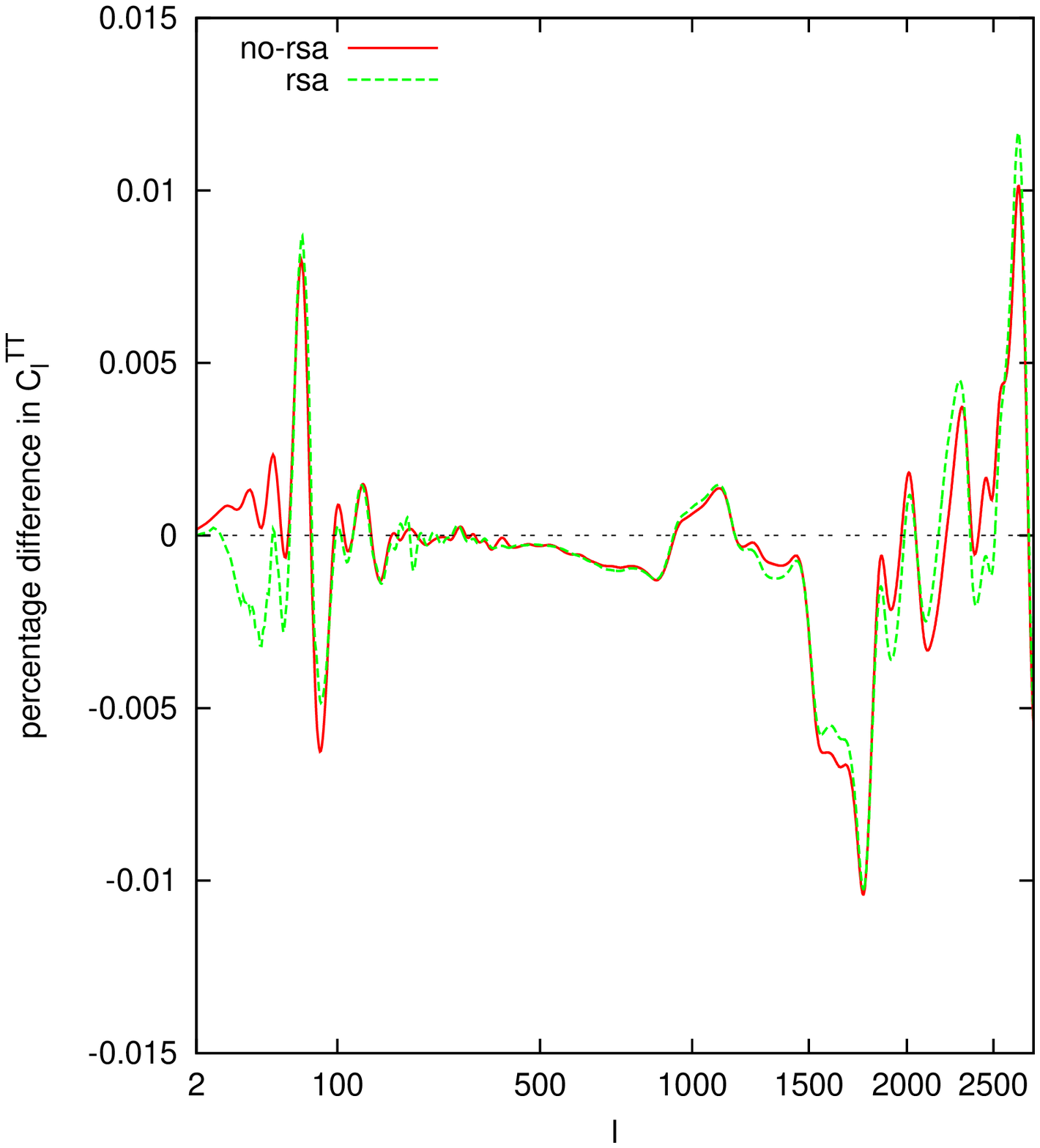}
\includegraphics[width=\wi]{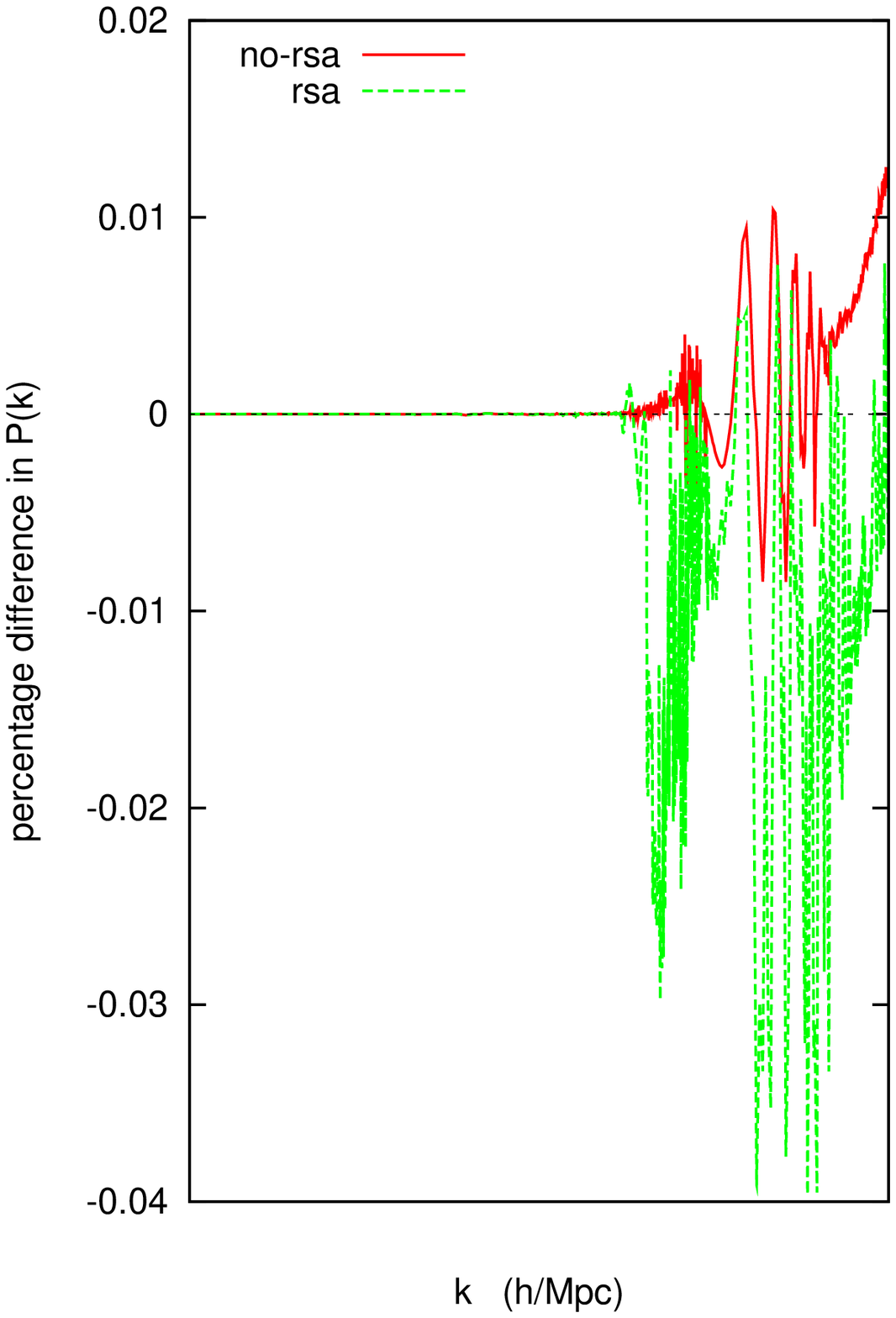}
\caption{\label{fa_cl1} Temperature and matter spectra for the runs
    {\tt rsa} and {\tt no-rsa}, normalized by reference
    spectra. Corresponding precision parameter settings are described
    in Table~\ref{fa_tab}. Using the RSA with such settings leads to
    equally accurate CMB spectra, and to a very small error on the
    matter power spectrum. The running time is however reduced
    considerably.}
%\end{figure}
}

\TABLE{
%\begin{table}
\begin{tabular}{|l|c|c|}
\hline
Runs with/without a RSA: & ~~~{\tt no-rsa}~~~ & ~~~{\tt rsa}~~~ \\
\hline
{\tt rk} & 55.8s & 33.6s  \\
{\tt ndf15} & 84.4s & 14.5s \\
\hline
\end{tabular}
\caption{\label{fa_time} Execution time of the perturbation module in
  seconds, with the precision parameters of the file {\tt
    cl\_permille.pre}, modified as described in Table \ref{fa_tab}.}
%\end{table}
}

In \CAMB{}, the RSA is somewhat cruder, since it neglects reionisation
and uses an explicit cosmology-dependent relation giving $h'$ an
$h''$ in terms of $\bar{\rho}_b$, $\bar{\rho}_{\cdm{}}$, $\delta_b$,
$\delta_{\cdm{}}$ and $k$, valid only deep in the matter-dominated
regime. We present in a companion paper the comparison between the
matter power spectrum $P(k)$ computed by \CAMB{} and \CLASS{}. In order to
get an accurate $P(k)$ with \CAMB{}, one is forced to deactivate the RSA
approximation (called {\tt late\_rad\_truncation}), precisely for the
above reasons. Our scheme does not lead to a significant error on the
$P(k)$, and is more model-independent: it involves only metric
perturbations and does not need to be modified in the presence of other
components playing a role during matter domination (e.g. with warm
dark matter or early dark energy).

%%%%%%%%%%%%%%%%%%%%%%%%%%%%%%%%%%%%
\section{Conclusions \label{sec_summary}}
%%%%%%%%%%%%%%%%%%%%%%%%%%%%%%%%%%%%

In Fig.~\ref{summary}, we summarise the different approximations used by the code
in the $(k, \tau)$ plane when computing the $C_l$'s up to 3000 and the
$P(k)$ up to $1 h$Mpc$^{-1}$. The figure corresponds to the
precision settings of the file {\tt cl\_3permille.pre}. Exact equations are used
only in the band corresponding to Hubble crossing for each mode, as well as in
the super-Hubble region with non-tightly-coupled photons, in which all
quantities evolve very slowly and integration is very fast.

With these approximations, for $\Lambda$CDM, the perturbation module
only spends a significant time in the region corresponding to Hubble
crossing for each mode, and to the sub-Hubble evolution before photon
decoupling. At early times, stiff equations are avoided thanks to the
Tight-Coupling Approximation.  Well-inside the Hubble radius and until
photon decoupling, the different modes oscillate and integration is
time-consuming: however, the number of equations is kept small (of the
order of 30 in total) thanks to the UFA.  After photon decoupling, the
code only needs to integrate over $4$ equations with very smooth
solutions, and the time spent by the code in the RSA region is
negligible.

Other approximation schemes can be introduced in more general
cosmological models. The case of massive neutrinos and non-cold dark
matter relics is discussed in a companion paper. More exotic cases may
require further approximations which can be introduced and discussed
case-by-case (\CLASS{} is coded in such way that introducing a new
approximation is as structured, codified and simple as introducing
new species \cite{class_general}). However, the fact that \CLASS{}
uses an original stiff integrator means that for several purposes (as
for the generalisation of TCA), new approximation schemes are not even
strictly necessary.

\FIGURE{
%\begin{figure}
%
\includegraphics[width=12cm]{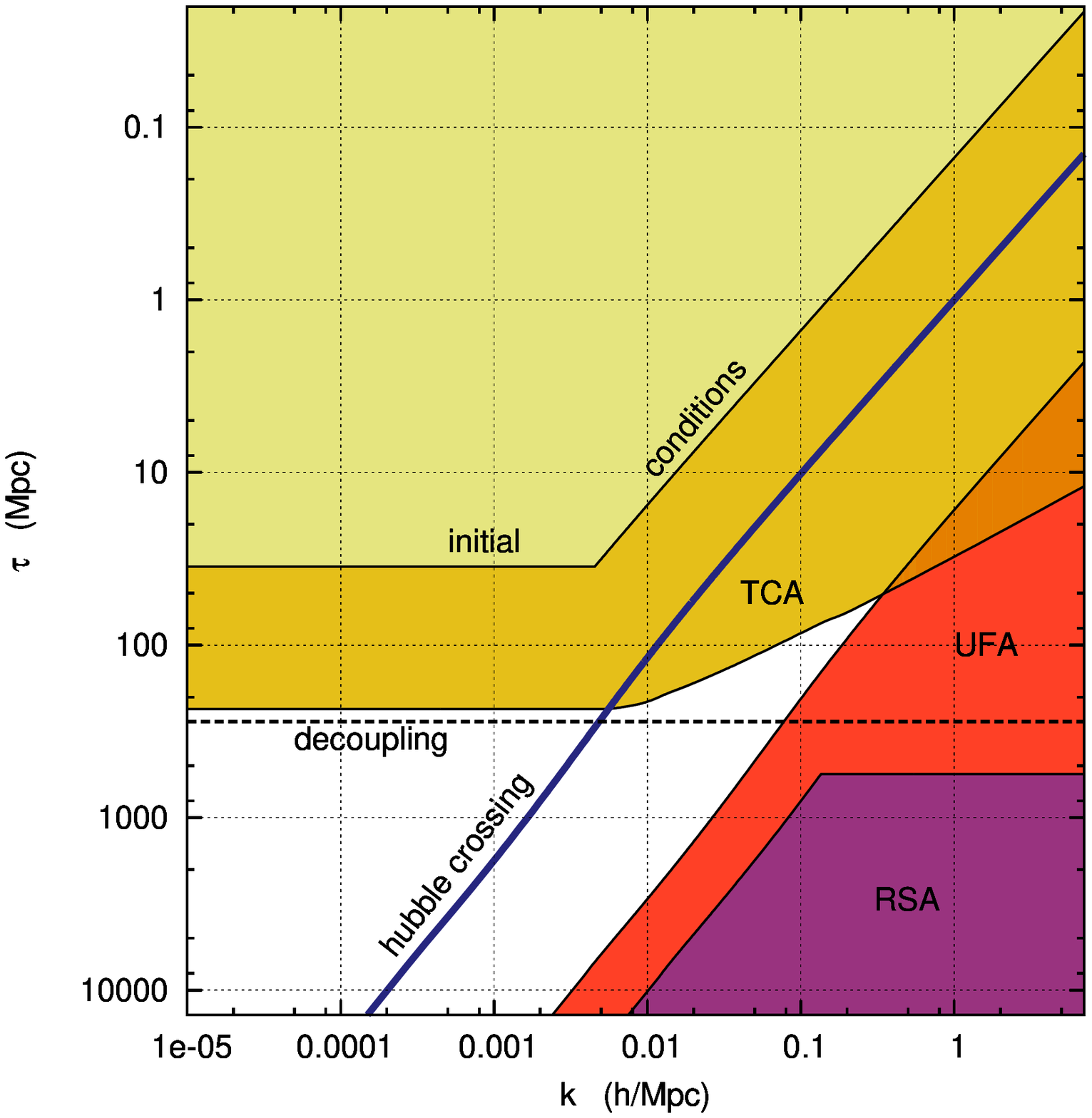}
%\
\caption{\label{summary} Summary of regions in $(k, \tau)$ space where
the various approximations are used. The precision settings are taken
from the {\tt cl\_3permille.pre} precision file which ensures a 0.3\%
precision on the $C_l$'s till $l=3000$. The full set of exact equations is
used only in the white region.}
%\end{figure}
}

\paragraph*{Acknowledgments}

This work was supported in part by the Swiss Science Foundation
(D.B.). Running with high-precision settings is only practical on
machines with many cores and large memory: we wish to thank
M. Shaposhnikov for providing us with a brand new 48-core PC at EPFL,
and the Institut d'Astrophysique de Paris for giving us access to the
horizon9 machine. Finally, we would like to thank A. Lewis for many
useful comments.

%%%%%%%%%%%%%%%%%%%%%%%%%%%

\appendix

%%%%%%%%%%%%%%%%%%%%%%%%%%
\section{Stiff integrator\label{sec_stiff}}
%%%%%%%%%%%%%%%%%%%%%%%%%%

The standard numerical method for solving Ordinary Differential
Equations (ODEs) is to use an adaptive step size Runge-Kutta
solver. While this method is fast and accurate in simple cases, it may
fail completely (or take a very large number of steps) when the
problem is stiff. Stiffness occurs when at least two times scales of
evolution in the problem differ substantially. A well known example is
the Boltzmann equation in cosmology. If a distribution is kept in
equilibrium by the (rapid) interaction with a background species, and
we are interested in the evolution of the distribution on cosmological
time-scales, a Runge-Kutta solver will oscillate around the
equilibrium solution by means of very small time steps related to the
time-scale of the interaction.

This problem exists, for instance, in the early universe where baryons
and photons are strongly coupled. One solution is to substitute the
equilibrium solution into the equations and thereby reducing the
system of equations and removing the stiffness: this is the basic
principle of the Tight Coupling Approximation (TCA) discussed in Sec.~\ref{sec_tca}. This approximation removes the stiffness of the Boltzmann
equation in the vanilla scenario.

However, stiffness may easily be reintroduced by trying to incorporate
new physics into the code.  The proper TCA approximation should then
be derived and implemented in the code, which would require a great
deal of familiarity with the code from the user. In \CLASS{}, however,
the user can just take advantage of the implemented stiff solver.

The {\tt ndf15} algorithm is a variable order (1-5) adaptive step size solver based on the Numerical Differentiation Formulas of order 1 to 5. The step size is adaptive but quasi-constant, meaning that the formulas used are based on a fixed step size. Each time the step size changes, the code will update the backward differences by interpolation to reflect this new step size. The
algorithm is described in~\cite{Shampine:1997}.

Whereas Runge-Kutta methods are explicit, meaning that the next step
can be computed directly from the previous step by elementary
operations, {\tt ndf15} is a fully implicit method. This means that at
each time step, we must solve a system of algebraic, possibly
non-linear equations, which is accomplished by Newton iteration. This
requires a numerical computation of the Jacobian and the solution of
systems of linear equations.  As it is evident, all this can make each
step very expensive, so a number of strategies must be implemented:

\begin{itemize}
	\item{Reusing Jacobians:} 
	
	The Jacobian usually changes more slowly than the solution itself, so an attempt is made to reuse the Jacobian - it will only be recomputed if Newton iteration is too slow.
	\item{LU decomposition:}
	 
	The same linear system must be solved repeatedly with different right hand sides, so we should of course store an LU decomposition. Because the system of equations can be large, $O(100)$, we need sparse matrix methods for this.
	\item{Backward Interpolation:}
	
	Since the method stores a matrix of backward differences, it is fast to infer the value and the derivatives of the solution at points before the current point by interpolation. We only need the values of some of the components to calculate the source functions, so we do not need to interpolate the rest of them.
\end{itemize}

When the number of equations is larger than about $10$, it is advantageous to use sparse matrix methods, and if the system is somewhat larger, the difference in execution time can differ by orders of magnitude. Usually sparse matrix methods are developed in order to save both time and memory, but for our purposes only execution time matters. We created a small sparse matrix package for our purpose based on~\cite{Davis:2006}. Some important features of this package are:

\begin{itemize}
	\item{Column Pre-ordering:}
	
          The matrices which appear are of the general form $I - cJ$
          where $I$ is the identity matrix, $J$ is the Jacobian and
          $c$ is some constant. Since the Jacobian is close to being
          structurally symmetric (if $y_i$ couples to $y_j$, it will
          often be the case that $y_j$ couples to $y_i$), it is
          advantageous to use the Approximate Minimum Degree column
          ordering of the matrix $C = J^T+J$. Using this pre-ordering
          reduces the number of non-zero elements in the corresponding
          L and U factors by a factor of a few, which leads to a
          sizable reduction in the time needed to solve the linear
          systems.
	\item{LU re-factorisation:}
	 
	If we have already calculated a LU factorization for some Jacobian, which is structurally identical to the current Jacobian, we can use information saved during the factorisation of the former to factorise the new matrix in a fraction of the time. Specifically, we store the pivot ordering and the \emph{reach} of all the sparse right hand sides used in forming the LU-decomposition.
	\item{Fast Jacobian Calculation:}
	
	If the same pattern of the Jacobian is found repeatedly, we can use this pattern to speed up the calculation. Taking advantage of the sparsity of the Jacobian, we can group the columns together and form the Jacobian using only a fraction of the usual $n$ function evaluations, $n$ being the number of equations.
\end{itemize}

The \CLASS{} user can choose to use the Runge Kutta or {\tt ndf15}
integrator by switching the precision parameter {\tt evolver} to
either {\tt rk} (=0) or {\tt ndf15} (=1, default setting). To
illustrate the power of {\tt ndf15} in stiff situations, {\tt ndf15}
was less than 10\% slower when the TCA was turned off as early as in
the reference run used in Fig.~\ref{tca_var}. As a comparison, the
standard Runge-Kutta integrator was more than 10000 times slower than
{\tt ndf15} for the same model. This particular run represents an extreme
case, but throughout this work we have presented various examples in which
the {\tt ndf15} performances are very good.

\section{Derivation of fluid equations for ultra-relativistic relics\label{appendix:fluid}}

The goal of this Appendix is to establish the validity of the approximate shear derivative equation~(\ref{ufa_class}), which allows to treat collisionless species as an imperfect fluid
governed by may less equations. In the future, our results could be used for computing higher order terms 
in (\ref{ufa_class}), or more generally for understanding various properties of the linear perturbations of ultra-relativistic species. A discussion similar in spirit was presented in~\cite{Shoji:2010hm} for massive neutrinos, although the goal of that paper was to introduce a sharp truncation at $l=3$, while we are searching  for a truncation scheme that would take into account the transfer of power to higher $l$'s.

\subsection{Formal solution}

Sticking to the notations of Ma \& Bertschinger, the perturbations of
ultra-relativstic species is described by a function $F({\bf
  k},\mu,\tau)$ obeying to the collisionless Boltzmann equation
\begin{equation}
\partial_\tau F({\bf k},\mu,\tau) +
i k \mu  F({\bf k},\mu,\tau) = S({\bf k},\mu,\tau)~,
\end{equation}
where $S$ stands for the gravitational source terms. The most general
formal solution can be written as
\begin{equation}
F({\bf k},\mu,\tau) = F^0({\bf k},\mu) e^{-i k \mu \tau}
+\int_0^\tau e^{-i k \mu (\tau-\tilde{\tau})}
S({\bf k},\mu,\tilde{\tau}) d \tilde{\tau}~.
\end{equation}
The initial function $F^0({\bf k},\mu)$ depends on the considered
gauge and type of initial conditions. For instance, in the synchronous
gauge, $F^0=0$ for the growing ADiabatic (AD), Baryon Isocurvature
(BI) and Cold Dark Isocurvature (CDI) modes; $F^0$ has a non-zero
monopole term for Neutrino Isocurvature Density (NID) initial
conditions; and a non-zero dipole term for Neutrino isocurvature
Velocity (NIV). These statements can be checked from ref.~\cite{Bucher:1999re}.
Hence, in all cases, we can write
\begin{equation}
F^0({\bf k},\mu)=C_{\rm NID}({\bf k})-ik\mu C_{\rm NIV}({\bf k})~.
\end{equation}
More fundamentally, the fact that $F^0$ only has monopole and dipole
contributions can be justified by the fact that neutrinos were
initially in thermal equilibrium, forming a fluid with no anisotropic
pressure or higher momenta. By causality, this remains true at any
time after decoupling on super-Hubble scales.  So, we are sure that
the above form of $F^0$ is completely universal.  In the synchronous
gauge, the source reads
\begin{equation}
S = - \frac{2}{3} h' - \frac{4}{3} (h'+6\eta')P_2(\mu)~.
\end{equation}
Thanks to a few integration by part, it is possible to absorb the
$\mu$ dependence, in order to be able to write this solution in
Legendre space.  The result is
\begin{eqnarray}
F({\bf k},\mu,\tau) &=& F^0({\bf k},\mu) e^{-i k \mu \tau}\\
&&+\frac{2}{k^2}\int_0^\tau 
e^{i k \mu (\tilde{\tau}-\tau)}
\left\{ 2 k^2 \eta'(\tilde{\tau}) + h'''(\tilde{\tau})+6\eta'''(\tilde{\tau}) \right\}
d\tilde{\tau} \\
&&-\frac{2}{k^2} \left\{ h''+6\eta''-i k \mu (h'+6\eta')\right\}\\
&&+ \frac{2}{k^2} e^{-i k \mu \tau} \left\{h''+6\eta''-i k \mu (h'+6\eta')\right\}_{\tau=0}~.
\end{eqnarray}
The last bracket contains the initial value of ($h'+6\eta')$
and $(h''+6\eta'')$. The former vanishes for all types of
initial conditions excepted NIV; the latter is non-zero for AD,
NID and NIV.  All initial condition terms in the first and last lines
can be grouped and represented by two
coefficients $\alpha$ and $\beta$:
\begin{eqnarray}
\alpha({\bf k})-i \mu\beta({\bf k})
&\equiv& F^0({\bf k},\mu) + \frac{2}{k^2}\left\{h''+6\eta''-i k \mu (h'+6\eta')\right\}_{\tau=0} \\
&=& \left\{ \delta_{\rm ur} + \frac{2}{k^2} (h''+6\eta'') \right\}_{\tau=0} 
- \frac{4 i \mu}{k} \left\{ \theta_{\rm ur}  + 
\frac{1}{2} (h'+6\eta')
\right\}_{\tau=0}~,
\end{eqnarray}
with e.g. $(\alpha, \beta) = (20/(15+4R_{\rm ur}),0)$ for adiabatic initial
conditions (as can be checked from \cite{Bucher:1999re}),
$R_{\rm ur}$ being the fractional contribution of
ultra-relativistic species to the background density.  
It is now easy
to expand the full solution in Legendre coefficients, using the
definition $F(\mu)=\sum_l (-i)^l (2l+1) F_l P_l(\mu)$ and the fact
that plane waves can be expanded in terms of spherical Bessel
functions:
\begin{eqnarray}
F_l({\bf k},\tau) &=& 
\alpha({\bf k}) j_l(k \tau) + \beta({\bf k}) j_l'(k \tau) \\
&&+ \frac{2}{k^2}
\int_0^\tau 
j_l\left( k (\tau-\tilde{\tau}) \right)
\left\{ 2 k^2 \eta'(\tilde{\tau}) + h'''(\tilde{\tau})+6\eta'''(\tilde{\tau}) \right\}
d\tilde{\tau} \\
&&-\frac{2}{k^2} \left\{ (h''+6\eta'') \delta_{l0} +\frac{k}{3} (h'+6\eta') \delta_{l1}\right\}~.
\label{eq:F_exact}
\end{eqnarray}
The terms in the first line show how initial conditions propagate to
later times, by just free-streaming. The other terms show how
perturbations adjust themselves to the power injected in the system at
any time by metric perturbations.

\subsection{Sub-Hubble approximation}

Well inside the Hubble radius, the above formal solution can be
approximated by a simpler expression. The results of this subsection
are never used in our UFA scheme or in {\rm CLASS}, but
we present them for completeness, and also because the approximation
performed in the next subsection will follow the same logic.

The second line of the solution contains a convolution between a
function which varies smoothly over a Hubble time (at least for $k
\tau\gg1$), and a Bessel function $j_l(x)$ with $x \equiv k
(\tau-\tilde{\tau})$ which oscillates over $\tau_k=1/k$. Bessel
functions $j_l(x)$ peak near $x_{\rm peak}=l+1/2$ (in fact this statement is
accurate only for very large $l$; for instance, $j_1(x)$ peaks near 
$x_{\rm peak}=2$
and $j_2(x)$ near $x_{\rm peak}=3.5$). The integral on $\tilde{\tau} \in [0,\tau]$
corresponds to $x \in [0, k \tau]$. The goal of this subsection is to
find an approximation for this convolution for small $l$ values.

As long as $k\tau \leq 1$, it is difficult to find a low-$l$ approximation
for the convolution; the result is a function oscillating over a
characteristic time $\tau_k=1/k$. In this regime, the integral brings
an extra oscillatory contribution to the term $\alpha({\bf k}) j_l(k
\tau) + \beta({\bf k}) j_l'(k \tau)$; this explains while around
Hubble crossing, the numerical solution for $F_l({\bf k},\tau)$
exhibits irregular oscillatory patterns, with very different peak
amplitude between two consecutive periods.

However, when $k\tau \gg 1$, the integral runs over a large range $x
\in [0, k \tau]$. For low $l$, this means that the convolution picks
up significant contributions only near $x=x_{\rm peak} \ll k \tau$.
Near this value, the slowly-varying argument can be approximated as a
constant, to be evaluated around $\tilde{\tau}=(k \tau-x_{\rm peak})/k$,
i.e. in very good approximation near $\tau$, since
$k \tau \gg x_{\rm peak}$. So, we can write:
\begin{eqnarray}
&&\int_0^\tau 
j_l\left( k (\tau-\tilde{\tau}) \right)
\left\{ 2 k^2 \eta'(\tilde{\tau}) + h'''(\tilde{\tau})+6\eta'''(\tilde{\tau}) \right\}
d\tilde{\tau} \nonumber \\
\longrightarrow&&
\left\{ 2 k^2 \eta'(\tau) + h'''(\tau)+6\eta'''(\tau) \right\}
\int_0^\tau 
j_l\left( k (\tau-\tilde{\tau}) \right) d\tilde{\tau}~.
\label{eq:convolution_approximation}
\end{eqnarray}
Finally, still in this limit $k\tau \gg 1$, the last integral can be approximated by
\begin{equation}
\frac{1}{k} \int_0^\infty
j_l\left( x \right) dx
= \frac{\sqrt{\pi} \Gamma(l/2+1/2)}{2k \Gamma(l/2+1)}
\end{equation}
A more accurate approximation scheme would lead to extra contributions involving time derivatives of the quantity between brackets in (\ref{eq:convolution_approximation}): for instance, the next order term would be of the type
\begin{equation}
\left\{ 2 k^2 \eta''(\tau) + h^{(4)}(\tau)+6\eta^{(4)}(\tau) \right\} \frac{\gamma}{k^2}
\end{equation}
with $\gamma$ being a coefficient of order one. However, since inside the Hubble scale metric perturbations vary over a
Hubble time $\tau \gg \tau_k$, we can keep only the leading 
source terms with the highest power of $k$:
\begin{eqnarray}
\delta_{\rm ur} = F_0 &=& 
\alpha j_l(k \tau) + \beta j_l'(k \tau) + \frac{4}{k}
\sqrt{\frac{\pi}{2}}
\eta'~, \\
\frac{4}{3k} \theta_{\rm ur} = F_1 &=& 
\alpha j_l(k \tau) + \beta j_l'(k \tau)
-\frac{2}{3k} h'~, \\
2 \sigma_{\rm ur} = F_2 &=& 
\alpha j_l(k \tau) + \beta j_l'(k \tau) + \frac{\pi}{k}\eta'~.
\end{eqnarray}
In the code, we do not use directly these asymptotic approximations,
and try instead to close the system of differential equations with a
trunction at order two, like for a viscous fluid.

\subsection{Exact truncation formula}

First, in order to manipulate more compact equations, we define:
\begin{equation}
G_l({\bf k},\tau) \equiv
F_l({\bf k},\tau)
-\frac{2}{k^2} \left\{ (h''+6\eta'') \delta_{l0} +\frac{k}{3} (h'+6\eta') \delta_{l1}\right\}~.
\end{equation}
Since $j_l'(x)=j_{l-1}(x)-\frac{l+1}{x}j_l(x)$ 
and $G_l$ contains a term $j_l(k \tau)$,
we suspect that it
obeys approximately to a similar relation. We use equation
(\ref{eq:F_exact}) for computing exactly the difference
$G_l'-kG_{l-1}+\frac{l+1}{\tau}G_l$, which would vanish if
only the term $\alpha j_l(k \tau)$ was contributing. We
use the identities
\begin{equation}
j_l''(x)=j_{l-1}'(x)-\frac{l+1}{x}j_l'(x)+\frac{l+1}{x^2}j_l(x)
\label{eq:identity}
\end{equation}
and $j_l(0)=\delta_{l0}$. We find
\begin{eqnarray}
G_l'-kG_{l-1}+\frac{l+1}{\tau}G_l 
&=&
\beta \frac{l+1}{k\tau^2}j_l(k \tau)
+\frac{2}{k^2} \left\{ 2 k^2 \eta'(\tau) + h'''(\tau)+6\eta'''(\tau) \right\} \delta_{l0} \nonumber \\
&&+ \frac{2}{k^2}
\int_0^\tau 
K(\tau,\tilde{\tau})
\left\{ 2 k^2 \eta'(\tilde{\tau}) + h'''(\tilde{\tau})+6\eta'''(\tilde{\tau}) \right\}
d\tilde{\tau}
\label{eq:exact_trunc}
\end{eqnarray}
where we defined
\begin{equation}
K(\tau,\tilde{\tau}) \equiv 
kj_l'\left( k (\tau-\tilde{\tau}) \right)
-kj_{l-1}\left( k (\tau-\tilde{\tau}) \right)
+\frac{l+1}{\tau}j_l\left( k (\tau-\tilde{\tau}) \right)
\end{equation}
This expression simplifies to
\begin{equation}
K(\tau,\tilde{\tau})
= -\frac{l+1}{\tau} \left(\frac{\tilde{\tau}}{\tau-\tilde{\tau}}\right)
j_l\left( k (\tau-\tilde{\tau})\right)
\end{equation}
This trunction scheme is not pratical in general. However,
the goal of the UFA is to find a way to close the system at low $l$
not at all times, but only deep inside the Hubble radius.

\subsection{Sub-Hubble truncation formula}

In the limit $k \tau \gg 1$, equation (\ref{eq:exact_trunc}) can be
simplified for two reasons. First, the Bessel function varies
over a time scale $\tau_k=1/k\ll\tau$, so
\begin{equation}
j_l'(k \tau) \gg \frac{j_l(k \tau)}{k\tau}
\end{equation}
Hence, in this limit, the last term in the
identity~(\ref{eq:identity}) can be omitted, which implies that the
first term (proportional to $\beta$) in eq.~(\ref{eq:exact_trunc}) is
always negligible, even in the presence of isocurvature modes.
Second, we can devise an approximation for the convolution, by
noticing once more that it involves metric perturbations wich vary
smoothly over a Hubble time (at least for $k \tau\gg1$), and the
quantity $\frac{k \tau-x}{x}j_l(x)$ with $x \equiv k
(\tau-\tilde{\tau})$ which oscillates over a period of order
$\tau_k=1/k$.

When $k\tau \gg 1$, the integral runs over a large range $x
\in [0, k \tau]$. For low $l$, this means that the convolution picks
up significant contributions only near $x=x_{\rm peak} \ll k \tau$,
while in this range $\frac{k \tau-x}{x}j_l(x) \simeq \frac{k \tau}{x}j_l(x)$.
Near $\tilde{\tau}=(k \tau-x_{\rm peak})/k\simeq \tau$, the slowly-varying
metric perturbations can be treated as a constant term.
So, we can write:
\begin{eqnarray}
&&\int_0^\tau
\left(\frac{\tilde{\tau}}{\tau-\tilde{\tau}}\right)
j_l\left( k (\tau-\tilde{\tau})\right)
\left\{ 2 k^2 \eta'(\tilde{\tau}) + h'''(\tilde{\tau})+6\eta'''(\tilde{\tau}) \right\}
 d \tilde{\tau}\nonumber \\
&\longrightarrow&
\left\{ 2 k^2 \eta'(\tau) + h'''(\tau)+6\eta'''(\tau) \right\}
\int_0^\tau
\left(\frac{\tilde{\tau}}{\tau-\tilde{\tau}}\right)
j_l\left( k (\tau-\tilde{\tau})\right) d \tilde{\tau}
\label{eq:convolution_approx2}
\end{eqnarray}
with
\begin{equation}
\int_0^\tau
\left(\frac{\tilde{\tau}}{\tau-\tilde{\tau}}\right)
j_l\left( k (\tau-\tilde{\tau})\right) d \tilde{\tau}
\simeq
\tau \int_0^\infty
x^{-1}
j_l\left(x\right) dx~.
\label{eq:approximate_integral}
\end{equation}
Let us estimate the error made in these approximations. Since we are inside the Hubble radius with smoothly varying metric perturbations, and since the integral in eq.~(\ref{eq:convolution_approx2})
is of order $\tau$, the leading term in eq.~(\ref{eq:convolution_approx2})
is of order $(k^2 \tau \eta')$. The other terms of order 
$(\tau \eta''')$ and $(\tau h''')$ can be neglected.
If instead of considering $\eta'(\tilde{\tau})$ as a constant we perform a Taylor expansion of this function around $\tau$, we find that the next order contribution to eq.~(\ref{eq:convolution_approx2})
is of order $(k \tau \eta'')$. Finally, an explicit calculation shows that the approximation performed in (\ref{eq:approximate_integral}) amounts in neglecting terms of order $k^{-1}$ with respect to terms of order $\tau$.
In summary, we obtain the following approximate truncation equation:
\begin{eqnarray}
G_l'-kG_{l-1}+\frac{l+1}{\tau}G_l 
&=&
\frac{2}{k^2} \left\{ 2 k^2 \eta'(\tau) + h'''(\tau)+6\eta'''(\tau) \right\} \delta_{l0} \nonumber \\
&&- 4(l+1)\eta'
\int_0^\infty
\frac{j_l(x)}{x} dx
+ \mathcal{O} \left( \frac{\eta''}{k}, \frac{\eta'}{k \tau} \right) ~.
\end{eqnarray}
For $l=2$, the integral is equal to -1/3, since
$(j_1(x)/x)'=j_2(x)/x$ and $\lim_{x\rightarrow0}[j_1(x)/x]=\frac{1}{3}$. So,
\begin{equation}
G_2'-kG_{1}+\frac{3}{\tau}G_2 
= 4 \eta'
+ \mathcal{O} \left( \frac{\eta''}{k}, \frac{\eta'}{k \tau} \right)~.
\end{equation}
Replacing the $G_l$'s by the appropriate momenta, we find:
\begin{equation}
\sigma_{\rm ur}' = - \frac{3}{\tau} \sigma _{\rm ur}
+ \frac{2}{3} \theta_{\rm ur} - \frac{1}{3} (h'+6\eta')
+ 2 \eta' + \mathcal{O} \left( \frac{\eta''}{k}, \frac{\eta'}{k \tau} \right)~.
\end{equation}
The last term $2 \eta'$ comes from our approximation for the convolution. Without a full treatement like the one presented here, one would miss this term and obtain the truncation formula called {\tt ufa\_mb} in the code, based on simply assuming $G_l \propto j_l(k\tau)$. However, with this extra contribution, the two terms in $\eta'$ cancel each other, and at leading order we end up with
\begin{equation}
\sigma_{\rm ur}' = - \frac{3}{\tau} \sigma _{\rm ur}
+ \frac{2}{3} \theta_{\rm ur} - \frac{1}{3} h'~,
\end{equation}
which is precisely what we call the {\tt ufa\_class} truncation scheme in
{\tt CLASS}.

\end{document}